%% file: main.tex
\newif\ifsigcomm
\sigcommfalse

\ifsigcomm
  \documentclass[sigconf,review,anonymous]{acmart}
\else
  \documentclass[sigconf,nonacm]{acmart}
\fi

\input{preamble}

\begin{document}

\title{Network-Enhanced Search: Restructuring Combinatorial Search to Exploit Communication}

\author{Paul Barford}
\email{pb@cs.wisc.edu}
\affiliation{\institution{University of Wisconsin - Madison}\country{Computer Sciences Department}}  % the 13 Nov version showed name and e-mail only; add an affiliation here if wanted

\input{sections/00-abstract}

\maketitle

\input{sections/01-introduction}
\input{sections/02-background}
\input{sections/03-model}
\input{sections/04-design}
\input{sections/05-implementation}
\input{sections/06-evaluation}
\input{sections/07-fabric}

\input{sections/08-discussion}
\input{sections/09-related}
\input{sections/10-conclusion}

\input{sections/11-acknowledgments}

\bibliographystyle{ACM-Reference-Format}
\bibliography{references}

\appendix
\input{sections/A-designs}
\input{sections/B-lessons}
\input{sections/C-negative}
\input{sections/D-theory}

\end{document}

%% file: preamble.tex
\usepackage{booktabs}
\usepackage{multirow}
\usepackage{algorithm}
\usepackage{algpseudocode}
\usepackage{xspace}
\usepackage{siunitx}
\ifsigcomm\else
  \renewcommand\footnotetextcopyrightpermission[1]{}
\fi

\newcommand{\sys}{NES\xspace}
\newcommand{\PAF}{\mathrm{PAF}}
\newcommand{\Nstar}{N^{*}}
\newcommand{\Z}{\mathbb{Z}}
\newcommand{\E}{\mathbb{E}}

\newcommand{\para}[1]{\smallskip\noindent\textbf{#1}\hspace{0.5em}}

%% file: sections/00-abstract.tex
%% ---------------------------------------------------------------- Abstract
%% Updated 19 Nov: FABRIC measurements and the eighteen new order-198 designs (sentences marked NEW below).
\begin{abstract}

AI-driven advances in automated reasoning have made large-scale distributed search a core scientific workload. Such
searches run as independent jobs that exchange almost nothing, leaving data-center networks idle and limiting how far a search can scale. We show that combinatorial search can be restructured so that communication raises the probability of success. Our system, Network-Enhanced Search (NES), partitions a search exactly into independent units and pools sampled candidate lists across nodes, so that $P$ nodes find about $P$ times more solutions for the same work, with every exchange's size and timing known before it starts. We report three key results. First, on the FABRIC testbed, pooling one join across 32 hosts reaches a first solution $5.9\times$ faster than one host and $1.6\times$ faster than the same hosts searching independently, and the network decides the gain, from $3.2\times$ on slow links to about $7\times$ on fast ones. Second, NES found {\em eighteen new D-optimal designs of order 198}, pairwise inequivalent and distinct from the one previously known, sixteen of them in a 384-core search coordinated over the network. Third, because demand is known in advance, applications can announce it to the network, and our measurement-parameterized model shows that at scale
pooled memory sets the largest solvable problem while reconfiguration delay, not bandwidth, limits accelerated nodes. NES lays the foundation for networks that are provisioned and controlled by the searches they serve, across the broad class of problems whose solutions are found by joining large candidate lists.
\end{abstract}

%% file: sections/01-introduction.tex
%% ---------------------------------------------------------------- 1 Introduction
\section{Introduction}
\label{sec:intro}

In 2026, AI laboratories began reporting mathematical and cryptographic discoveries made by machines. OpenAI announced advances in mathematics and theoretical computer science, including explicit constructions that settle open questions or refute conjectures~\cite{r55}, and later reported claimed proofs of several longstanding problems under expert review~\cite{r56}. Anthropic reported improvements on the lower bound for Riemann zeta zeros on the critical line, new attacks on cryptographic primitives, and the discovery of a Hadamard matrix of order 668, open since 2004~\cite{r6,r39}. These breakthroughs share a common pattern: search-and-verify workloads that generate vast numbers of candidates, reject most through inexpensive tests, and rigorously verify the few survivors. As these searches expand from a single model's context window to distributed populations of agents and machines, the organization and exchange of intermediate results become fundamentally systems and networking problems.

Data-center networks have repeatedly evolved to match their workloads. For example, Hedera corrected misplaced large flows~\cite{r3}; Helios and c-Through added optical circuits for persistent bulk traffic~\cite{r25,r73}; coflow schedulers optimized data-parallel transfers around collective deadlines~\cite{r16,r19}; and modern ML fabrics co-design topology with training parallelism~\cite{r38,r41,r74}. These designs differ in what they know and when, but they share one premise:  communication is overhead. The computation is fixed, it emits traffic, and the network's job is to make that traffic cost as little as possible.

This paper studies a search-and-verify workload for which that premise fails. The workload consists of three stages: nodes generate large sets of candidate pieces; pairs or quadruples of pieces are combined, with only those whose statistics sum exactly to a target surviving; the rare survivors are then verified exactly, yielding certificates that anyone can check. We focus on classic combinatorial search problems, including the existence of Hadamard matrices and D-optimal designs, whose defining constraints are precisely such autocorrelation sums. Interest in these searches has recently accelerated: program search has produced new extremal constructions~\cite{r64}, and the first D-optimal design of order 198 was discovered in October 2026 after roughly 104 hours on 32 cores~\cite{r27,r28}. Because these searches naturally execute on clusters, they are compelling targets for fast, reconfigurable optical fabrics. We therefore begin by asking whether solver telemetry can be used to steer such networks.

Today's search codes leave the network little to do. Before building anything, we analyzed 63 network traffic traces of six workload classes, including parallel SAT with clause sharing, island-model sequence search, meet-in-the-middle joins, state-vector simulation, HPC traces, and AI training traces (\S\ref{sec:background}, Appendix~\ref{app:negative}). None satisfied even basic requirements for communication intensity, reconfiguration opportunity, and predictability. Clause sharing consumed only about 0.004\% of link capacity, while heavy traffic was largely stationary and served as well by a fixed topology chosen per job as by a reconfiguration oracle. These codes move little data because they were designed for networks that are slow and expensive, avoiding communication by design. This suggests that the opportunity lies in restructuring the search itself, not in serving existing traffic more efficiently.

\para{Insight: restructure the search so that communication amplifies success probability.} A solution is assembled
from several pieces, and the search tries combinations of candidate pieces until one satisfies the defining condition.
Because every candidate can combine with every other, the chance of success grows much faster than the number of
candidates. Pooling candidate lists across $P$ nodes therefore lets each node's pieces combine with everyone else's, and for the same total work finds about $P$ times more solutions than $P$ nodes searching independently. We realize this structure through exact coarse-to-fine stages, a {\em compression tower} (our term for the multi-level compression of~\cite{r46,r48}), and join sampled candidate lists at the dominant stage using a linear hash of the summed statistic. In this design, communication is not a by-product of computation; it directly increases the probability that a search round succeeds. We call the resulting system NES (Network-Enhanced Search, \S\ref{sec:impl}).

\para{What the network is for: coordination and memory, not only bandwidth.} We identify two regimes. When a stage splits exactly into independent pieces of work ({\em e.g.,} the order-198 D-optimal search), nothing on the critical path crosses the network: communication is under 0.1\% of time to solution, and nodes need only receive shared starting data and learn when to stop. When a stage combines pooled candidates ({\em e.g.,} Hadamard search) the network matters in a different way. Pooling across hosts was up to $5.9\times$ faster than one host and $1.6\times$ faster than the same hosts searching independently; once the exchange was on the critical path, slowing the links from 25 to 0.5\,Gb/s more than doubled time to solution and erased pooling's advantage. At larger scale, pooling also lets many nodes hold one candidate list too large for any single server, and our model, parameterized by measurements, predicts that this pooled memory, rather than compute or link speed, sets the largest problem that can be solved (sequences about 15\% longer at 1,024 nodes). The network is thus a coordination plane and a memory-expansion engine for search, and what decides which fabric works best is how the exchange is scheduled, including reconfiguration delay and whether demand is announced ahead of time.

\para{A new area: network-enhanced search.} We argue that this workload opens a research area: search restructured to
exploit the network and, ultimately, networks provisioned and controlled by the searches they serve, joining networking, systems and algorithms. It differs from the workloads data-center networks are designed for in four ways. (1)~\emph{The output is a certificate.} Finding the object is hard and checking it is cheap, so the metric is time to discovery, not throughput or job completion time, and correctness rests on exact verification and honest accounting of what was searched. (2)~\emph{Communication raises the probability of success.} Pooled joins find about $P$ times more solutions per unit of work, so traffic is a source of gain rather than overhead, and the key decision each round is how large to make the pooled list, which sets both the success probability and the size of the next exchange. (3)~\emph{Demand is announced, not estimated.} The application knows the size of its next exchange (a fixed number of bytes per candidate times the list size $N$ it has just chosen) and when it will start, before the round’s computation ends. (4)~\emph{Coordination can shrink to almost nothing.} When the work splits into independent pieces, nodes share only a small amount of starting data and a stop signal; when candidates are pooled, they share only the candidate lists themselves, whose size is known in advance.

\para{More combinations, not better luck.} A natural way to speed up a random search is to restart runs that seem unlucky ({\em i.e.,} not finding a solution) or to pick promising starting points. Neither works here: the
chance of finding a solution is the same in every round, however long a run has gone without one, and starting
points are close to interchangeable, so choosing them helps little (1.12--1.19$\times$, and only in a narrow
regime). What does shorten the search is raising the number of candidate combinations tested per unit of work,
either by pooling lists across nodes or by choosing which combinations to run~\cite{r27}. 

\para{The same structure decides what to compute and what to exchange.} Combining two candidate lists of sizes
$N_A$ and $N_B$ tests $N_AN_B$ combinations for $N_A+N_B$ work. Pooling exploits this across nodes; an earlier
search~\cite{r27} exploits it within a node, running first the combinations that offer the most candidates per unit
of work. We find that solutions appear in proportion to this count (slope 0.96--0.97), which turns that heuristic
into a predictive model: on a 64-thread server it yields 6.1--6.3$\times$ more expected designs per CPU-hour than
checking all work in order. Selecting work this way needs almost no coordination: the best cutoff is the same
everywhere, so nodes that rank only their own work come within 3\% of a global ranking. Coordinating an exact
search therefore means sharing one number, not a work queue.

\para{Contributions.}
\begin{itemize}
\item A new workload class for network design, search-and-verify workloads in which communication raises the
  probability of success, so that pooled work across $P$ nodes finds about $P$ times more per unit of work; and a measurement study of 63 network traffic traces of existing search, HPC and AI-training workloads, in which communication did not raise the probability of success, motivating the creation of this property by restructuring the algorithm (\S\ref{sec:background}).
\item A model of communication-amplified search: how hits and work grow with list size, the critical list size
  beyond which pooling stops paying, and measurements showing that waits are memoryless and starting points nearly interchangeable, with supporting theory (\S\ref{sec:model}, Appendix~\ref{app:theory}).
\item A partitioned search design: compression towers, an exact join of sampled candidate lists, small independent
  final checks, and list sizes chosen each round that can be announced to the fabric before the exchange
  (\S\ref{sec:design}).
\item \textbf{Eighteen new D-optimal designs of order 198.} The restructured search found eighteen D-optimal designs of
  order 198, pairwise inequivalent and inequivalent to the one previously published design~\cite{r27}, at the sequence level (cyclic shifts, negations and multipliers) and as $198\times198$ matrices (all nineteen 4-row inner-product profiles differ, even allowing transposition). Two were found on one server, the second after 11.8 hours of wall-clock time; sixteen by a 384-core search whose nodes were coordinated over a network, ten of them in one 48-hour stretch on 256 cores. Every design is verified exactly (\S\ref{sec:eval-198}, \S\ref{sec:fabric},
  Appendix~\ref{app:designs}).
\item Evaluation on two problem families with independently verified solutions: the solvable Hadamard length rises
  from 72 to 120 on one server (known orders), with $v=112$ (order 452) in a median of about 10 minutes per run (6/6) and $v=120$ (order 484) in about 6.2 hours (5/6, 98.8 CPU-hours per solution), where the previous version solved none of four runs in 36 hours each; D-optimal designs rise from 57 to 75 on single cores (known orders); growth per unit length falls from $\ge2.16\times$ to $1.28\times$; pooling gives $P^{0.99}$ in a controlled emulation; and server-scale engineering makes the exact order-198 check 3.83$\times$ faster on identical work (\S\ref{sec:eval}).
\item Measurements on the FABRIC testbed and a measurement-parameterized model showing what the network contributes: pooling across hosts is up to 5.9$\times$ faster than one host and 1.6$\times$ faster than independent hosts; once the exchange is on the critical path, link rate changes time to solution by up to 2.2$\times$, so the speedup over one host ranges from 3.2$\times$ on slow links to about 7$\times$ on fast ones; and at scale, pooled memory rather than compute or bandwidth sets the largest solvable problem (\S\ref{sec:eval-network},
  \S\ref{sec:fabric}).
\end{itemize}
The negative results that shaped the design (restart policies, projection methods, cross-parent joins, symmetry
restrictions, screening, and a pilot at an open order) are collected in Appendix~\ref{app:negative}.

%% file: sections/02-background.tex
%% ---------------------------------------------------------------- 2 Background and Motivation
\section{Background and Motivation}
\label{sec:background}

\subsection{Search for extremal objects}

\begin{figure*}[t]
  \centering
  \includegraphics[width=\textwidth]{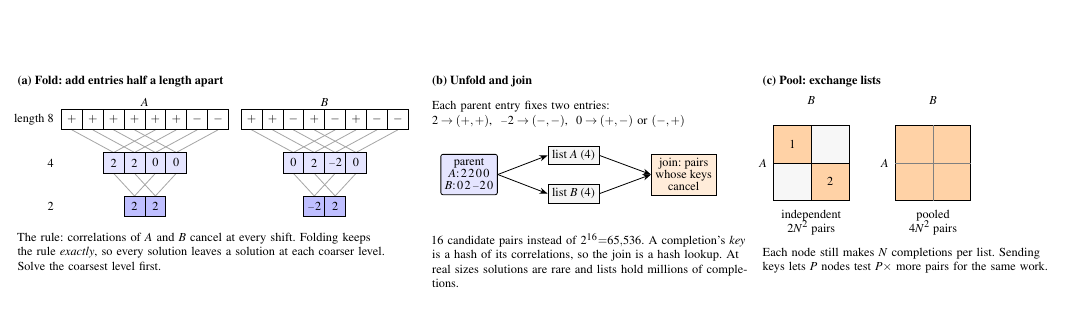}
  \caption{A compression tower on a toy problem: two $\pm1$ sequences of length 8 whose periodic correlations cancel
  at every nonzero shift. (a) Folding gives exact smaller problems. (b) A coarse solution (parent) fixes pairs of
  entries and leaves short lists of completions, which a hash join matches. (c) Exchanging key lists across nodes
  multiplies the candidate pairs per unit of work. Formal treatment in Appendix~\ref{app:theory}.}
  \label{fig:tower}
\end{figure*}

Hadamard matrices and D-optimal designs are related families of $\pm1$ matrices with applications in coding, communication, and experimental design. A Hadamard matrix has mutually orthogonal rows, while a D-optimal design maximizes determinant. In the constructions studied here, a Hadamard matrix is built from four $\pm1$ sequences and a two-circulant D-optimal design from two, whose periodic autocorrelations (PAFs) must sum to $-4$ and $2$, respectively, at every nonzero shift. These are exact search problems: a candidate either satisfies all constraints or it does not, and any solution can be verified independently ($HH^{\top}=nI$ or by checking the determinant). The search space contains $2^{4v}$ or $2^{2v}$ candidates; the remaining open instances have largely resisted algebraic constructions~\cite{r54,r68}, and finding an exact D-optimal design is NP-hard.

Figure~\ref{fig:tower} illustrates the method on a toy problem: find two $\pm1$ sequences $A$ and $B$ of length 8 whose agreements cancel at every nonzero shift. The agreement at shift $s$ is the number of matches minus mismatches between a sequence and a copy rotated by $s$. For example, $A=(+,+,+,+,+,+,-,-)$ has agreements $4,0,0,0,0,0,4$, while $B=(+,+,-,+,-,+,-,-)$ has $-4,0,0,0,0,0,-4$, making $(A,B)$ a solution (a periodic Golay pair). Hadamard and D-optimal searches impose the same type of constraint on four or two sequences at much larger lengths. Rather than examine all $2^{16}$ pairs, we proceed in four steps. {\em Fold.} Adding entries half a length apart reduces length 8 to 4, then to 2. Every solution remains a solution after folding, creating a hierarchy of exact smaller problems, a compression tower. {\em Unfold.} Solving the smallest level and expanding upward generates candidate completions. In the figure, one length-4 parent yields 4 completions of $A$ and 4 of $B$: 16 candidate pairs instead of 65,536, of which 8 are solutions. Each parent is an independent unit of work. {\em Join.} Hashing correlation values allows matching completions to be found by lookup, so two lists of $N$ candidates test $N^2$ pairs with roughly $O(N)$ work. {\em Pool.} At realistic sizes, solutions are rare and nodes sample only $N$ candidates per list. Two nodes searching independently test $2N^2$ pairs; exchanging lists tests $4N^2$ for the same work, and $P$ nodes gain nearly a factor of $P$ in general. Exchanged data raises the chance of success; the rest of the paper builds on this property.

\para{How such searches run today.} Practical searches combine algebraic restrictions with enumeration. Compression, the folding above, preserves the defining condition exactly at a shorter length~\cite{r53} and underlies the recent D-optimal records~\cite{r54,r27}; applied in stages, it makes exhaustive searches feasible for related sequences~\cite{r48,r46}. Our compression towers are this multi-level structure; what we change is how the dominant level is searched and what is exchanged there. Today these searches run as fully independent jobs: workers take disjoint parts of the space and report back. Parallel SAT, cube-and-conquer and cryptanalytic collision search divide or share work in similar ways~\cite{r9,r66,r31,r32,r71,r72}.

\subsection{Beyond Hadamard: where the pattern applies}
\label{sec:generality}

Hadamard matrices and D-optimal designs are our test cases, but our design applies to a broad class of search problems. What the network exploits is a pattern that many search and cryptanalytic problems share, and whether a problem has it is easy to check. A problem fits when it has three properties.

\para{(P1) An exact reduction.} The problem reduces at low computational cost to a smaller version of itself that keeps every solution in reduced form, so the small
version can be solved first, and each of its solutions confines the rest of the search to an independent unit of
work. Examples: folding a sequence, reducing modulo a prime or subgroup, projecting onto a few coordinates, and
fixing the top bits of a sum.

\para{(P2) An additive key.} A candidate solution is assembled from parts, and whether it works can be decided by
comparing short keys computed from each part separately, because the combined key is the sum (or XOR) of the parts'
keys. A join on keys then examines all combinations of parts while computing each part only once. This is the
meet-in-the-middle and generalized-birthday structure~\cite{r67,r72}.

\para{(P3) Sampled lists that reward pooling.} The full lists are too large to enumerate, so each node (computer or search process) samples $N$ candidates per list. With $k$ lists, combinations grow as $N^k$ but work far more slowly, so pooling $P$ nodes' samples finds about $P$ times more solutions than $P$ separate nodes for the same work(\S\ref{sec:model}). This property turns communication into success probability.

\para{Where it holds.} Table~\ref{tab:generality} lists problems with the same structure. Sequence-design searches (Williamson and Turyn-type sequences, periodic Golay and Legendre pairs) use the same folds and keys as our test cases; our Williamson join already gains $7.6$--$24\times$ from shared lists (Table~\ref{tab:sharing}). Meet-in-the-middle algorithms for subset sum and knapsack join on partial sums modulo $2^b$~\cite{r76,r77}; code decoding joins on syndrome fragments~\cite{r79}; Wagner's $k$-tree and parallel collision search join on hash bits~\cite{r71,r72}; and lattice sieving joins vector pairs whose sum is short~\cite{r78}. In each case, the dominant step joins sampled lists, success grows superlinearly with list size, and each candidate contributes only a short key. The networking consequences of \S\ref{sec:model} therefore carry over: pooled lists require an all-gather (or, at large $P$, a list-partitioning shuffle) once per round, exchange volume is determined by a preselected list size, and communication can be scheduled in advance. Section~\ref{sec:discussion} discusses how to validate these properties for each problem.

\begin{table}[t]
  \caption{Problems with the same structure. ``Reduction'': the exact reduction to a smaller version (P1). ``Key'': what the join compares (P2). ``Lists'': $k$ in $N^k$ (P3). Measured rows are from this paper; the others follow from the cited algorithms' structure and are not measured here.}
  \label{tab:generality}
  \small
  \begin{tabular}{@{}p{2.25cm}p{1.65cm}p{1.75cm}p{0.6cm}p{0.65cm}@{}}
    \toprule
    Problem & Reduction (P1) & Key (P2) & Lists & Here \\
    \midrule
    Hadamard (4 seqs) & fold & PAF vector & 4 & meas. \\
    D-optimal (2 seqs) & fold, double & PAF vector & 2 & meas. \\
    Williamson, Golay, Legendre pairs~\cite{r46,r48} & fold & PAF vector & 2--4 & Tab.~\ref{tab:sharing} \\
    Subset sum / knapsack~\cite{r76,r77} & sum mod $2^b$ & partial sums & 4--8 & -- \\
    Code decoding (ISD)~\cite{r79} & bits of syndrome & syndrome & 2--4 & -- \\
    Generalized birthday, hash collisions~\cite{r71,r72} & top bits & hash XOR & 2--$2^d$ & -- \\
    Lattice sieving~\cite{r78} & bucketing & vector sum & 2--3 & -- \\
    \bottomrule
  \end{tabular}
\end{table}

\para{Where it does not.} The pattern breaks down when constraints cannot be decomposed into additive per-part keys, as in aperiodic problems such as LABS whose correlations do not fold exactly; when shared information must be integrated into solver state before becoming useful, as in SAT clause sharing~\cite{r9,r66}; or when the reduction partitions the search space so finely that candidates cannot match across nodes. The last case is not a failure. Such problems, including our order-198 search, reduce to many small independent jobs, the second traffic class of \S\ref{sec:design}. Distinguishing these regimes, and identifying which applies to a given problem, is one contribution of this paper. Proof-of-work mining ({\em e.g.,} Bitcoin) provides a clean negative control: each candidate is tested independently against a random-oracle hash, leaving no fold or join. Hits and work therefore scale linearly, and by the luck ceiling (Appendix~\ref{app:theory}) pooling provides no gain ($P^{0}=1$). In that regime, the network can help only by rapidly propagating stop signals, which we measure in \S\ref{sec:fabric}.

\subsection{Where existing network designs fall short}

Network designs for computational workloads fall into five groups (Table~\ref{tab:position}, \S\ref{sec:related}).
Demand-inferred designs observe traffic and react~\cite{r3,r10,r14,r25,r73}; demand-oblivious fabrics cycle through matchings regardless of demand~\cite{r4,r8,r50,r51}; application-exposed schedulers let the application describe transfers it has already decided to make~\cite{r15,r16,r18,r19}; co-designed ML fabrics choose topology and parallelism together~\cite{r38,r41,r47,r74}; and in-network computing moves part of the computation into switches~\cite{r37,r45,r65}. All treat the bytes as fixed by the application and minimize their cost. Our experiments
run on a standard packet network, which is enough to show the gain, but none of these designs exploits it: none
restructures an algorithm so that {\em sending more data} makes it succeed more often, or takes as input a decision
made before the exchange (the list size) that sets both the traffic and the probability of a result.

\begin{table}[t]
  \caption{Where \sys sits. ``Demand'': how the network learns the traffic. ``Role'': what communication is to the
  computation.}
  \label{tab:position}
  \small
  \begin{tabular}{@{}p{2.15cm}p{1.45cm}p{1.35cm}p{2.05cm}@{}}
    \toprule
    Group (examples) & Demand & Role & Network decision \\
    \midrule
    Demand-inferred~\cite{r3,r10,r25,r73} & estimated from counters & overhead & flow/circuit placement \\
    Demand-oblivious~\cite{r4,r8,r50,r51} & ignored & overhead & fixed schedule \\
    App.-exposed~\cite{r15,r18,r19} & declared, fixed & overhead & ordering, endpoints \\
    Co-designed ML~\cite{r38,r41,r47,r74} & from job graph & overhead & topology, parallelism \\
    In-network~\cite{r37,r45,r65} & fixed & offloaded work & what runs in switch \\
    \midrule
    \sys & announced, before compute ends & source of gain ($\approx P\times$) & list size $N$, pooled vs.\ independent \\
    \bottomrule
  \end{tabular}
\end{table}

\subsection{A negative motivation: today's search traffic}

We asked whether search workloads could benefit from a reconfigurable network steered by solver telemetry. Three conditions must hold: the network must matter (a 400,Gb/s link speeds a run by at least $1.25\times$), reconfiguration must pay (an oracle reconfiguring every 10,\textmu s beats the best fixed topology by at least 15\%), and solver telemetry must predict traffic shifts better than past traffic (by at least 0.10 PR-AUC). We tested 63 traces from six workload classes: SAT clause sharing, LABS island search~\cite{r57}, Williamson meet-in-the-middle, state-vector simulation, HPC~\cite{r52,r69}, and AI training. None satisfied all three criteria (Appendix~\ref{app:negative}). Search workloads move little data because their algorithms avoid communication; where traffic is heavy (AI training), it is stable and already well served by fixed co-designed topologies~\cite{r38,r74}.

\begin{table}[t]
  \caption{Network demand with the network removed (bytes per node per second of compute) and load $\rho$ on a
  400\,Gb/s link. \sys rows are from the model of \S\ref{sec:eval-network}.}
  \label{tab:demand}
  \small
  \begin{tabular}{@{}p{4.0cm}p{1.75cm}p{1.6cm}@{}}
    \toprule
    Workload (traces/runs) & demand/node & $\rho$ @ 400G \\
    \midrule
    HPC mini-apps (DUMPI, ATLAHS; 64--1024 ranks) & 9--177\,MB/s$^{\dagger}$ & 0.0002--0.0035 \\
    SAT clause sharing (Mallob, 64--128 ranks) & 1.4--1.9\,MB/s & $\approx4\cdot10^{-5}$ \\
    AI training (Llama 7B, MoE; GOAL) & 48--56\,GB/s & 0.96--1.11 \\
    \sys join, $v=96$, $P=64$, CPU nodes & $\approx0.3$\,MB/s & $<10^{-5}$ \\
    \sys join, same, nodes $10^3\times$ faster & $\approx0.24$\,GB/s & $\approx0.005$ \\
    \bottomrule
    \multicolumn{3}{@{}l}{\footnotesize $^{\dagger}$ per MPI rank}
  \end{tabular}
\end{table}

Table~\ref{tab:demand} shows that, unlike AI training, search workloads consume only a tiny fraction of available network capacity: HPC mini-apps need 9--177,MB/s per rank, parallel SAT about 2,MB/s per rank, and our pooled join well under 1,MB/s per node (only about 0.5\% of a 400,Gb/s link even with nodes $10^3\times$ faster), making time-to-result gained per byte communicated and per microsecond of reconfiguration a more meaningful metric than utilization.
%That is the design point we explore.

\begin{table}[t]
  \caption{Sharing results in existing search classes (compute scaled $100\times$ to model accelerated nodes):
  useful, but small and saturating.}
  \label{tab:sharing}
  \small
  \begin{tabular}{@{}p{2.5cm}p{1.75cm}p{1.6cm}p{1.35cm}@{}}
    \toprule
    Algorithm & Scale & Gain & Data shared \\
    \midrule
    Williamson MITM join & $n=31$, 16/64 nodes & 7.6$\times$ / 24$\times$ & 94--98\,MB \\
    Ramsey canonical table & 4/16/64 workers & 0.97/1.2/2.3$\times$ & 0.12--0.16\,GB \\
    LABS bound broadcast & $N=32,34$; 16/64 & 2.7--4.7$\times$ & 0.2--23\,KB \\
    \bottomrule
  \end{tabular}
\end{table}

Communication helps search only when recipients can exploit what they receive. Sharing results reduced time to result by 7.6--24$\times$ for a Williamson join, 2.3$\times$ for a Ramsey canonical-form table, and 2.7--4.7$\times$ for LABS bound broadcasts (Table~\ref{tab:sharing}), but gains saturated above 25--100,Gb/s because the shared objects are tiny relative to the work they eliminate. By contrast, heavier clause sharing in cube-and-conquer CaDiCaL moved 17--29$\times$ more data and slowed five of six runs by 16--72\%, because every imported clause must be propagated. Pattern databases, read in place, continued improving with bandwidth (1.7--2.5$\times$ from 100,Gb/s to 1.6,Tb/s). The design lesson is simple: traffic should be targeted, compact, and immediately usable. %A received key in our join is one hash insert or probe.

%% file: sections/03-model.tex
%% ---------------------------------------------------------------- 3 Communication-Amplified Search
\section{Communication-Amplified Search}
\label{sec:model}

Our central claim is that search can be restructured around exact joins so that pooled data directly increases the probability of success, and that the gain is bounded in ways that inform network design.

\para{The model.}   In this paper, we develop a model that predicts time to solution for a pooled search from a few measured quantities: how hits grow with list size ($\lambda\propto N^{\alpha}$), how work grows ($\propto N^{\beta}$), and the cost of each round's exchange. The model's three results are the pooling gain $P^{\alpha/\beta-1}$, the critical list size $\Nstar$, and the
expected time to a first solution (Eq.~\ref{eq:ET}); \S\ref{sec:eval-network} adds the network term for each fabric (Eq.~\ref{eq:tnet}).

\para{Hits and work.} Consider a join that draws \(N\) candidates from each of \(k\) sequences and reports tuples whose keys cancel exactly. Each draw-and-join is a \emph{round}; the search repeats rounds until a hit is found. With independent sampling, the expected number of hits per round is $\lambda(N)=cN^{k}$, where $c$ is
the probability that a random tuple is a solution (by linearity of expectation, whatever the sampling distribution). Work per round grows as $N^{\beta}$, where $\beta$ is the work exponent: $\beta=1$ for a two-list hash join and $\beta\approx2$ for a four-list join that enumerates pairs. (\S\ref{sec:eval}).

\para{Pooling.} Pooling candidate lists from \(P\) nodes into one larger list yields about \(P\) times more solutions than searching independently, while exchanging only each candidate's short key. Let \(\alpha\) denote the measured hit exponent (\(\lambda\propto N^{\alpha}\), nominally \(k\)). \(P\) independent nodes with lists of size \(N_0\) find \(PcN_0^{\alpha}\) hits for \(PN_0^{\beta}\) work. A pooled list of size \(N=P^{1/\beta}N_0\) costs the same work but finds \(cP^{\alpha/\beta}N_0^{\alpha}\) hits, a gain of \(P^{\alpha/\beta-1}\). For both our problems, \(\alpha/\beta\approx2\) (4/2 for Hadamard and 2/1 for D-optimal), giving an approximately linear \((\approx P)\) benefit. Because a four-list key is the sum of per-list keys, nodes exchange only key lists (12\,B per candidate), not pairs: each node gathers all lists and enumerates the pairs whose combined key hashes to its assigned partition.

\para{One hit is enough.} Define \(\Nstar\) by \(\lambda(\Nstar)\approx1\). Well below \(\Nstar\), most rounds fail; above it, additional hits have little value. Assuming Poisson-distributed hits, a round succeeds with probability \(1-e^{-\lambda(N)}\), giving 
\begin{equation} \E[T]=\frac{t_{\mathrm{comp}}(N)/P+t_{\mathrm{net}}(N,P)}{1-e^{-\lambda(N)}}, \label{eq:ET} 
\end{equation} 
where \(t_{\mathrm{comp}}(N)\) and \(t_{\mathrm{net}}(N,P)\) are the computation and exchange times per round. Pooling therefore reduces search time while \(\lambda(N_0)\ll1\), but provides little benefit once a pooled round is likely to succeed. Because \(\Nstar\) grows rapidly with problem size (about \(3\times\) for every \(+8\) in Hadamard length \(v\)), outpacing the memory and compute of a single node, pooling's main practical role is enabling lists to reach \(\Nstar\).

\para{Choosing the list size.} For \(\lambda(N)\ll1\), Eq.~\ref{eq:ET} becomes \(\E[T]\approx (t_{\mathrm{comp}}(N)/P+t_{\mathrm{net}})/(cN^{\alpha})\), which scales as \(N^{\beta-\alpha}\) when computation dominates. Since \(\alpha>\beta\), larger lists reduce time to solution until \(\lambda\) approaches 1, placing the optimum near \(\Nstar\). Pooling multiplies hits per round by \(P^{\alpha/\beta}\), giving \(\lambda=P^{\alpha/\beta}\lambda(N_0)\), and reduces search time by about \(P^{\alpha/\beta-1}\) (roughly \(P\)) only while this remains \(\lesssim1\), i.e., while individual nodes are far below \(\Nstar\). Beyond that point, pooling yields more hits per round but little improvement in time to the first solution. Because \(\lambda\) is directly observable, a controller can identify this operating point within a few rounds.

\para{Memorylessness: luck cannot be exploited.} If some runs were predictably luckier than others, search could beat pooling by restarting poor runs or choosing promising starting points. We tested this directly (Appendix~\ref{app:negative}) on roughly 5,500 traced runs. Waiting times at frontier-scale list sizes are exponential (CV 0.96--1.00), parent success is independent of prior attempts, Hadamard parent features are uninformative (AUC 0.49--0.53), D-optimal fertility is only weakly predictable (AUC 0.61--0.67, too weak to justify testing), early-run predictors~\cite{r34,r40} achieve just 43--55\% accuracy against a 50\% baseline, restart policies never outperform completion, and solutions are uniformly distributed across starting points (dispersion index 1.00--1.10). In such searches, the only way to reduce \(\E[T]\) is to increase \(\lambda\) per unit of work, either by choosing more productive joins (\S\ref{sec:eval-198}) or by pooling lists, which increases \(\lambda\) much faster than work. Pooling is the network's contribution.

\para{What this asks of a network.} The model turns into three requirements. \emph{R1, aggregation:} the fabric must
let $P$ nodes act as one list of size $P^{1/\beta}N_0$, which needs an all-gather of key lists per round, or at large $P$ a partitioned exchange that pools memory across nodes. \emph{R2, timeliness:} each round's exchange is on the critical path; once compute is accelerated, its latency, including any circuit reconfiguration, adds directly to every round. \emph{R3, fan-out:} exact final checks produce many tiny independent jobs whose inputs are a few bytes. R1 and R2 suit circuits scheduled from announced demand (\S\ref{sec:eval-network}); R3 suits packets.

\para{Theory of towers and joins (Appendix~\ref{app:theory}).} (T1) \emph{Folds are exact, but tower depth is limited.} Folding by \(m\) sums entries whose positions agree modulo \(v/m\), preserving the spectrum exactly at frequencies that are multiples of \(m\), so the condition holds exactly at length \(v/m\); folds compose~\cite{r48}.  A tower is therefore a divisor chain of \(v\), chosen by a small dynamic program. Frequencies coprime to \(v\) (30 conjugate pairs at \(v=99\)) are lost by every fold and must all be matched in the final join, so deeper towers reduce intermediate cost but not final difficulty, explaining their limited benefit (\S\ref{sec:eval}). (T2) \emph{The best join depends on cost ratios and \(P\).} With candidate cost \(c_q\), pair cost \(c_p\), and \(S\) key values, the optimal list size is \((c_pS/c_q)^{1/3}\), and Wagner's merge~\cite{r72} is optimal when \(c_q=c_p\). Expensive candidate generation, as in our problems, favors a full join on one node; at large \(P\), an all-gather's per-candidate cost no longer shrinks with \(P\), and list partitioning becomes preferable. (T3) \emph{There is a ceiling on candidate selection.} With uniformly distributed keys, \(N\) candidates per list contain \(N^k/S\) expected solutions regardless of how they are chosen. Beyond pooling, gains therefore require cheaper candidates or non-uniform solution density; relabeling keys cannot help.

%% file: sections/04-design.tex
%% ---------------------------------------------------------------- 4 Design
\section{Design}
\label{sec:design}

\sys restructures search in four steps and exposes the result to the network as two traffic classes and one control
decision (Figure~\ref{fig:tower}; Figure~\ref{fig:pipeline} in Appendix~\ref{app:theory} shows the full pipeline).

\subsection{Splitting search into independent units}

The toy example of \S\ref{sec:background} folded a length-8 problem to length 2, solved it there, and unfolded the result. At realistic sizes, the same idea does more than shrink the search: it decomposes one large problem into many small, independent units of work that can be distributed, ranked, and scheduled.  We next describe how towers are built and how these units are defined.

\para{Building a tower.} Folding by a divisor \(m\) of \(v\), \(y_j=\sum_{t<m}x_{j+t v/m}\), maps the problem exactly to length \(v/m\) (target \(-4m\) for Hadamard, \(2m\) for D-optimal), so repeated folds form a tower of ever smaller exact problems. We solve the coarsest level first and then unfold: each coarse value constrains the sum of \(m\) finer values, confining refinement to a small subspace. A coarse solution (a \emph{parent}) thus becomes a unit of work. Each level is solved by the method best suited to its size: exhaustive search at the top, local search in the middle, and joins where most of the cost resides. When \(v\) admits only one useful fold (\(v=166=2\cdot83\) for order 668), the coarse level itself becomes the bottleneck and is solved with the same join machinery.

\para{Double compression: exact units of work.} When \(v=pq\) with \(\gcd(p,q)=1\), a sequence can be viewed as a \(p\times q\) array, whose row and column sums are obtained by folding by \(q\) and \(p\), respectively. Fixing multiple compressions can leave subproblems small enough to enumerate exactly, giving each parent an exact yes-or-no test for extendability, itself decomposed into independent jobs. For order 198 (\(v=99=9\cdot11\)), a unit fixes a pair of length-33 compressions (fold by 3) and a pair of length-9 compressions (row sums). Together these determine the length-11 column sums and decompose each sequence into three independent \(3\times11\) blocks. Testing a unit then enumerates all consistent full-sequence pairs, so the unit contains a design iff the test finds one. The same construction applies to Hadamard lengths with coprime factorizations (\(80=16\cdot5\), \(96=32\cdot3\), \(104=8\cdot13\)), where it replaces middle-level local search with a small exact enumeration per parent (\S\ref{sec:eval-solvable}).

\subsection{The sampled exact join}

At each unfold step we draw \(N\) completions per sequence, retain only those satisfying the spectral bound (an exact filter), and search for tuples whose agreement vectors meet the target. A linear 64-bit hash of a sequence's agreements, $h(x)=\sum_s\PAF_x(s)R_s\bmod2^{64}$, makes the key of a pair the sum of two keys, so the Hadamard join is a four-sum over integers ($h_A+h_B=T-h_C-h_D$) and the D-optimal join a two-sum; every key match is rechecked exactly. To limit memory, pairs are partitioned into \(B\) hash buckets based on the high bits of \(h_A+h_B\) and joined independently in cache, so the full pair table is never materialized. Buckets are also the unit of distribution: with \(P\) nodes, each node owns a subset of buckets, receives all key lists, and enumerates only its assigned pairs.

\subsection{Exact final checks as small jobs}

\para{Engineering the exact check.} In double compression, the final level is an exact yes-or-no test whose cost is dominated by combining precomputed block lists and rejecting combinations that violate the spectral bound. We improve its speed and robustness (\S\ref{sec:eval-198}) in four ways: \emph{exact filtering,} using a fast prescreen followed by exact verification of every survivor; \emph{symmetry elimination,} testing rotationally or otherwise equivalent candidates only once; \emph{coordinator-free scheduling,} deriving each unit from a fixed global numbering so late arrivals and restarts never duplicate work; and \emph{exact accounting,} counting a unit only after complete evaluation and recording precisely which subunits finish when a worker stops early.

\begin{algorithm}[t]
  \caption{One pooled join round at node $i$ of $P$ (four-list Hadamard join).}
  \label{alg:round}
  \small
  \begin{algorithmic}[1]
    \State $N\gets$ controller's list size for this round; $(\ell_i,u_i)\gets$ node $i$'s bucket range of $h_A+h_B$
    \State \textsc{announce}(peers $=\{1..P\}$, bytes $=48N(1-1/P)$, start $\approx$ now $+\hat t_{\mathrm{gen}}$)
      \Comment{fabric sets up first matching}
    \State draw $N/P$ spectrally feasible completions per sequence $A,B,C,D$ for the current parent; key each by $h(\cdot)$
    \State all-gather the four key lists (key, id: 12\,B each) \Comment{traffic class (a)}
    \For{each $AB$ pair with $h_A+h_B\in[\ell_i,u_i)$, radix-partitioned into cache-sized buckets}
      \State probe $CD$ pairs with $h_C+h_D=T-(h_A+h_B)$; for each match, fetch 4 ids and recheck PAF exactly
    \EndFor
    \State report hits, duplicates and time shares; on a verified solution, stop all nodes
    \State controller: $N\gets$ update$(N,\text{hits})$ aiming at $\lambda(N)\approx1$; keep the parent if no hit
      (parents are interchangeable)
  \end{algorithmic}
\end{algorithm}

\begin{figure}[t]
  \centering
  \includegraphics[width=\columnwidth]{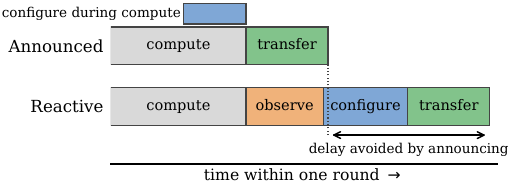}
  \caption{One join round, schematic. With demand announced, the exchange's size and start time are known before computation ends, so a fabric that must reconfigure can do so during compute and the transfer starts immediately. A fabric that reacts to traffic must first observe the exchange, then reconfigure.}
  \label{fig:announced}
\end{figure}

\subsection{Traffic classes and announced demand}

The design produces two kinds of traffic. (a)~\emph{One bulk exchange per join round}: every node gathers all key
lists, a fixed number of bytes per candidate times the list size $N$. (b)~\emph{Many small independent jobs}: the exact tests, whose inputs are a few bytes and whose outputs are
almost always empty. Class (a) is on the critical path, but its size and timing are known before it starts, which a reconfigurable fabric can exploit.  Class (b) fits an ordinary packet network. 

\para{The controller.} A small controller runs alongside the solver and makes two decisions each round. First, it
sets the list size: from the round's hit count it estimates $\hat\lambda$ and sets $N\gets N(1/\hat\lambda)^{1/\alpha}$,
changing $N$ by at most a factor of 4 per round, aiming at about one hit per round (Eq.~\ref{eq:ET}); in simulation
on measured hit curves this settles within a few rounds (\S\ref{sec:eval-network}). It also chooses between pooled
and independent search. Second, it \emph{announces demand}: because it has just fixed $N$, it can tell the network
the participants, the volume per node and the start time of the next exchange before the round's computation ends.

\para{What the network can do with an announcement.} Existing fabrics either infer demand from traffic
counters~\cite{r3,r25,r73} or cycle through connection patterns regardless of demand~\cite{r8,r51}. With demand
announced, a fabric that must reconfigure can set up its first configuration while nodes are still computing and
schedule the rest in advance; an all-gather among $P$ nodes with $d$ circuits each needs $\lceil(P-1)/d\rceil$ such
configurations. Algorithm~\ref{alg:round} gives one pooled round, and Figure~\ref{fig:announced} contrasts it with a
fabric that reacts to observed traffic. The small exact jobs bypass this entirely and are dealt over the packet
network from the fixed numbering of units. When several searches share a fabric, each can adjust the size of its
announced exchanges, which makes the joint schedule a coflow problem~\cite{r19} with adjustable volumes
(\S\ref{sec:discussion}).

%\para{Decisions the controller does not need.} Our measurements rule out two decisions one might expect a controller to make: abandoning runs that look unlucky (waits are memoryless, \S\ref{sec:model}), and sharing records of parents already shown to hold no solution (only 0.1--2.3\% of 9,930 Hadamard parents recurred across 33 runs).

%% file: sections/05-implementation.tex
%% ---------------------------------------------------------------- 5 Implementation
%% Updated 17 Nov: final-join kernel (had_dc6w), distributed pooled join over TCP, FABRIC deployment.
\section{The NES Kit}
\label{sec:impl}

NES (Network-Enhanced Search) performs all computations reported in this paper. The code-base was written for this study and reuses no existing solver, though it implements established techniques: sequence compression~\cite{r53,r46,r48},
Gale--Ryser margin feasibility, and multi-list joins~\cite{r67,r72}. The solvers are written in C.  Scheduling, coordination, verification and analysis are
written in Python~3. NES has four main functions. \emph{Hadamard search} builds towers with any sequence of
folds, chooses local search or a join at each level, and runs the bucketed join of \S\ref{sec:design}, on one node or
pooled across many. \emph{D-optimal search} adds coarse-level matching and double compression; for order 198
($v=99=9\cdot11$) it searches exhaustive libraries of length-9 compressions (54,065 and 52,070 sequences) and an
exhaustive middle level at length 33, then runs each parent's exact test over three $3\times11$ blocks per sequence, split into independent jobs across cores or nodes. \emph{Verification} checks every reported solution independently, by an exact determinant for D-optimal designs and $HH^{\top}=nI$ for Hadamard matrices; we rebuilt and verified the published order-668 matrix the same way. \emph{Scheduling} runs unattended on a 64-thread server (two 16-core Xeon Gold 5218 CPUs, 188\,GB RAM) and across testbed hosts, handling multi-core jobs, wall-clock limits, memory reservations, crash-safe accounting across code versions, and automatic verification.

\para{The final join.} The Hadamard solver's final join has four optimizations. It walks the lists in key order,
which finds the same hits at 1.4--2.6$\times$ lower cost per pair examined. It stores each candidate in a compact
8-byte form, so lists four times larger fit in memory. It drops parents that have too few candidates to fill a list,
which caused every time-out in the previous version. And it lets 16 parallel workers build one parent's lists together and then split its buckets, with the first verified hit stopping the others. With these options off, its output is identical to the previous version's on every seed tested.

\para{Distributed execution.} The pooled join runs as one process per node, connected by TCP over the testbed's
network. Each round, nodes gather everyone's key lists (12--16\,B per candidate), join the buckets they own, and stop
at the first verified hit, which the finding node broadcasts. For the order-198 search, one coordinator per site sends every node the library of coarse solutions, deals out disjoint blocks of units, collects a short record
($\sim$100\,B) for each finished job along with any solutions, re-verifies each design independently, checks it
against all known designs, and broadcasts a stop signal. If the coordinator is unreachable, nodes keep searching and
resend what it missed when it returns.

\para{Correctness.} Every reported design is verified exactly. Before use, each optimized solver is compared output for output with its previous version on the same seeds: the order-198 solvers agreed on tens of thousands of units at $v=45$ and 63 and on full $v=99$ units; the exact test split into parallel jobs matched the serial test on every parent tried; and the pooled join's hits across 4 nodes on one parent equal a single process's exactly (677 of 677). An independent review of NES with memory and undefined-behavior checkers found the cache bug reported in
Appendix~\ref{sec:eval-lessons}, whose effect was then audited from logs. Before any run, on every machine, known-answer tests re-find the published order-198 design and both of our server designs through the full production path.

%% file: sections/06-evaluation.tex
%% ---------------------------------------------------------------- 6 Evaluation
%% Updated 17 Nov: kit v5c frontier results (6.1), designs 3-4 and campaign totals (6.3, Table 8), FABRIC lesson (6.4),
%% pointer to measured FABRIC results (6.5).
\section{Evaluation}
\label{sec:eval}

We evaluate five questions: Does restructuring expand what is solvable (\S\ref{sec:eval-solvable})? Does pooling
deliver the gains the model predicts, and at what network cost (\S\ref{sec:eval-pooling})? Can an engineered exact
test, work chosen by join volume and a network-coordinated search find new order-198 designs
(\S\ref{sec:eval-198})? And what must the network provide (\S\ref{sec:eval-network})? Unless stated otherwise, experiments in this section run on a 64-thread server (two 16-core Intel Xeon Gold 5218 CPUs, base clock 2.3\,GHz, 188\,GB RAM), and times are single-core CPU seconds to an independently verified solution (medians; unsolved runs count as above their limit).

\subsection{Restructuring changes what is solvable}
\label{sec:eval-solvable}

\begin{figure*}[t]
  \centering
  \includegraphics[width=\textwidth]{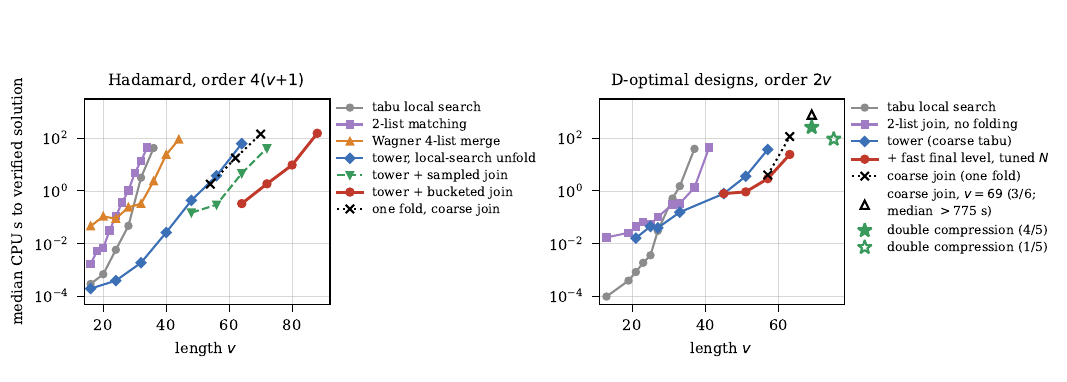}
  \caption{Median CPU seconds to a verified solution vs.\ length. Left: Hadamard (bordered four-sequence
  construction); the tower uses three folds of 2 except the one-fold series. Right: D-optimal designs; the tower folds
  by 3 (by 5 at $v=25$). Open markers are censored or single-success points.}
  \label{fig:growth}
\end{figure*}

\para{Growth rates.} To test whether restructuring changes how cost grows with problem size, we measure time to a
solution as the sequence length $v$ increases (3--8 seeds per size, Figure~\ref{fig:growth}). The tower with a
bucketed join cuts the growth per unit length of Hadamard search from at least $2.16\times$ for tabu search~\cite{r81}, a standard local search, to $1.28\times$, so every 24 units of length it gains more than five orders of magnitude. Each step of the restructuring contributes: a plain join, Wagner's four-list merge~\cite{r72} and a tower that unfolds by local search fall in between. D-optimal search shows the same pattern, from $2.03\times$ for tabu search, which stalls at $v=41$, to $1.25\times$ for the tower, which solved every run up to $v=57$.

\begin{table}[t]
  \caption{Hadamard critical path (three folds of 2). Local-search tower: total time and final-level share; joins:
  median time with lists near $\Nstar$.}
  \label{tab:critical}
  \small
  \resizebox{\columnwidth}{!}{\begin{tabular}{@{}lrrrrr@{}}
    \toprule
    $v$ (order) & LS tower & final share & sampled join & bucketed join & memory \\
    \midrule
    48 (196) & 0.15\,s & 97.9\% & 0.35\,s & -- & -- \\
    56 (228) & 0.30\,s & 96.2\% & 3.9\,s & -- & -- \\
    64 (260) & 4.6\,s & 87.8\% & 64\,s & 0.33\,s & -- \\
    72 (292) & 39--44\,s & -- & -- & 1.85\,s & -- \\
    80 (324) & none in 15\,min & -- & -- & 9.5\,s & 0.53\,GB \\
    88 (356) & -- & -- & -- & 47--257\,s & 1.6\,GB \\
    \bottomrule
  \end{tabular}}
\end{table}

\begin{table}[t]
  \caption{D-optimal designs: each fix targets the level that dominated before it. Median CPU s (solved/runs).}
  \label{tab:dopt}
  \small
  \begin{tabular}{@{}lrrrr@{}}
    \toprule
    $v$ (order) & tower & + fast final, $N$ & + coarse join & + double compr. \\
    \midrule
    51 (102) & 3.6 (8/8) & 0.92 (8/8) & -- & -- \\
    57 (114) & 37 (8/8) & 2.9 (8/8) & 4.0 (8/8) & -- \\
    63 (126) & none, 240\,s & 24 (8/8) & 114 (8/8) & -- \\
    69 (138) & -- & 0/3 & 3/6 (126--775) & 257 (4/5) \\
    75 (150) & -- & 0/3 & 0/3 & 92 (1/5) \\
    \bottomrule
  \end{tabular}
\end{table}

\para{The critical path moves, and each fix follows it.} We instrumented every stage of both searches and removed
bottlenecks as they appeared (Tables~\ref{tab:critical} and~\ref{tab:dopt}). In Hadamard search, the final level
consumes 88--98\% of runtime and failed because local search could not find completions that existed: every tested
parent was completable. Replacing local search with a sampled join, then adding linear hashing and bucketing, reduced \(v=64\) from 64\,s to 0.33\,s (\(\sim200\times\)) and made \(v=80\) solvable in 9.5\,s using 0.53\,GB, where no solution had been found in 15 minutes. When only a single fold is available, the bottleneck shifts to the coarse level, where the same join is far faster and solves lengths local search could not. In D-optimal search, the
bottleneck moved from candidate generation to the coarse level and finally to parent quality: only about 2\% of
parents extend to a solution, and a sampled round often misses even those. Double compression resolves this by giving each parent an exact test, which solved \(v=75\).  No single change was enough: each fix moved the bottleneck, and following it made lengths that no earlier method
could solve tractable.

\para{The frontier at scale.} To test whether the tower's low growth rate holds at the largest sizes one core can
reach, we ran it on Hadamard lengths up to $v=96$ (order 388). It solved every run at $v=80$ and 88 (medians 29\,s and 245\,s) and 28 of 36 at $v=96$ within 2.2\,h, with median time growing about $8\times$ per $+8$ in length, as
$1.28\times$ per unit predicts. Two further findings point to what comes next. A deeper tower did not help, as the
depth limit of \S\ref{sec:model} predicts, and run times, though they varied more than $30\times$, followed an
exponential wait rather than a heavy tail (Figure~\ref{fig:evidence}c, Table~\ref{tab:v96}), so restarts cannot help either. These orders were constructed before~\cite{r54,r68}, and we claim no mathematical records here; our times, measured as single-core CPU seconds to a verified solution on stated hardware, offer a benchmark against which future methods can be compared.The point is that the growth rate holds at scale and neither a deeper tower nor luck moves the frontier further; pooling across nodes does.

\begin{table}[t]
  \caption{Hadamard $v=96$ (order 388) on the server: 12 seeds per configuration, single core, 2-hour limit. Mean
  wait: total CPU including censored runs divided by solutions (exponential estimate).}
  \label{tab:v96}
  \small
  \begin{tabular}{@{}lrrr@{}}
    \toprule
    Tower (folds; method per level) & solved & median & mean wait \\
    \midrule
    3 levels (2,2,2; join at last level) & 11/12 & 1,845\,s & 3.1\,ks \\
    4 levels (2,2,2,2; join at last two) & 9/12 & 1,458\,s & 4.5\,ks \\
    4 levels (2,2,2,2; join at last level) & 8/12 & 2,932\,s & 6.8\,ks \\
    \bottomrule
  \end{tabular}
\end{table}

\para{Pushing the single-server frontier to $v=120$.} Once local search was gone from the final level, we analyzed how far one server could go by removing each remaining bottleneck in turn. At $v=96$ the middle level had become the bottleneck. Two coprime folds turn each parent there into a few small units that can be enumerated exactly, which made the middle level about $350\times$ faster (from about a minute to under a second) and left the final join with more than 99.9\% of the time. We then engineered that join (\S\ref{sec:impl}), $5.3\times$ faster per core, and split each parent's join across 16 workers. The frontier moved from $v=96$ to $v=120$ (order 484): the server solved
$v=104$ in under a minute, $v=112$ in about 10 minutes and $v=120$ in about 6.2 hours (5 of 6 runs), where the
previous version solved none of four $v=120$ runs in 36 hours each (a bound until a paired comparison completes).
These orders are known, and every solution was verified independently; the gain is in time to a solution. What
remains is the final join alone, the step whose success grows with list size and therefore with pooling.

\para{Sampling where solutions live.} Solutions are not spread evenly over a parent's candidates. Their components are spectrally flatter than typical candidates, so keeping only the flattest 30--50\%, which we call \emph{tilting}, raises the odds that a sampled candidate belongs to a solution. \emph{Tilting finds 5.3--10.3$\times$ more verified hits per CPU in isolated joins at $v=72$}, and in full tower runs it brings the first solution about $2$--$3\times$ sooner at $v=80$ and 96. \emph{But it competes with pooling:} tilting shrinks the pool of completions a parent offers, so at the list sizes of a pooled round its gain vanishes and some tilted lists cannot fill. Tilting therefore helps independent nodes most, and combining it with pooling means choosing the tilt and the pool size together (Appendix~\ref{app:negative}). It is the one technique that we found that departs from the uniform model of \S\ref{sec:model}, the kind of gain that learned samplers could extend (\S\ref{sec:discussion}).

\subsection{Pooling multiplies the chance of success}
\label{sec:eval-pooling}

\begin{figure*}[t]
  \centering
  \includegraphics[width=\textwidth]{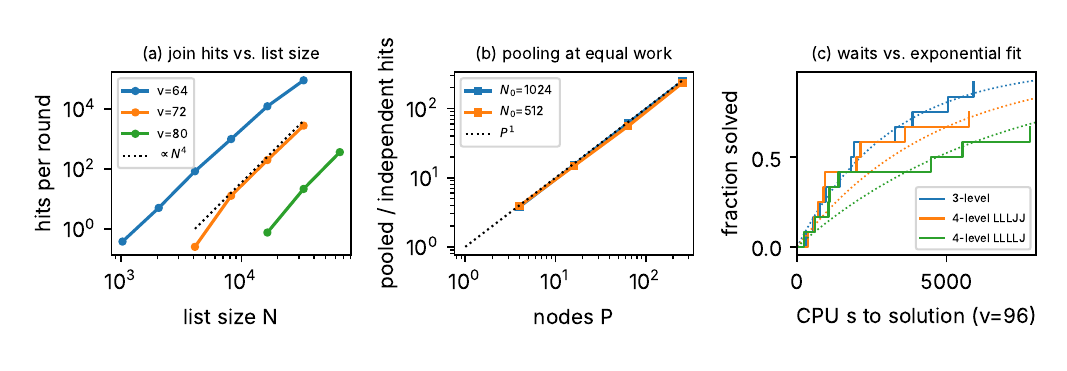}
    \caption{Measurements testing the model of \S\ref{sec:model}, all from single-core runs on the 64-thread server.  (a)~Hits per round of the Hadamard join, measured by repeatedly joining lists drawn from fixed parents at $v=64$, 72 and 80: hits grow as about $N^{4}$ with list size $N$ (dotted line), while work grows only about as $N^{2}$, the gap that makes larger lists pay. (b)~Pooled against independent hits per node at equal work per node, from an emulation of $P$ nodes on the server in which each emulated node owns a range of buckets and its own CPU clock: pooling finds about $P$ times more (dotted line: $P^{1}$; measured $P^{0.99}$). (c)~Fraction of Hadamard $v=96$ runs solved against CPU time, for three towers (12 seeds each, 2-hour limit; L and J mark levels solved by local search or join), with exponential fits (dotted): waits are memoryless, so restarting a run cannot shorten the search, and a deeper tower does not help.}
  \label{fig:evidence}
\end{figure*}

\para{Objective and result.} This subsection tests the two predictions of the model in \S\ref{sec:model}: that hits
grow faster than work, and that pooling $P$ nodes therefore finds about $P$ times more. We emulate $P$ nodes on one server. Each emulated node generates its own share of the candidate lists, owns a range of buckets in the pooled join, and keeps its own CPU clock; the emulator always advances the node whose clock is furthest behind, and the first verified hit by node clock wins, so results match what $P$ nodes running in parallel would see, minus the cost of the exchange. \emph{Both predictions hold.} For fixed Hadamard parents, hits per round scale as \(N^{3.8\text{--}4.1}\) while cost scales as \(N^{1.7\text{--}1.95}\), so hits per CPU-second grow roughly as \(N^2\) (Figure~\ref{fig:evidence}a). The D-optimal join shows the same pattern (\(\alpha\approx2\), \(\beta\approx1\)). Holding work per node fixed, pooling increases hits per node by \(P^{0.99}\) from 4 to 256 nodes (Figure~\ref{fig:evidence}b), a result confirmed on FABRIC (\S\ref{sec:fabric}). Exchanging key lists rather than pairs keeps communication modest: at \(P=64\), a round transfers about 99\,MB in total versus 34\,GB for pairs. Each key is consumed immediately by a single insert or probe, allowing joins to overlap communication. The critical list size \(\Nstar\) roughly triples for every \(+8\) increase in length, and because parents are interchangeable, time to solution is governed by list size rather than parent choice. For one-fold D-optimal instances, pooling first becomes important at the coarse level, which requires roughly 4\,M candidates per sequence per round, beyond what a single node can generate quickly.

\para{The limits of pooling.} Pooling shortens the search only while a single node's list is far below $\Nstar$.
Where one node is already close, pooling 16 nodes gave $14.8\times$ more hits but only $1.6\times$ shorter time,
because a single pooled round already succeeds. The regime where pooling pays most, $\lambda(N_0)\ll1$, is exactly the
regime of open instances, where $\Nstar$ outgrows one node's memory and compute.

\subsection{Finding new order-198 designs}
\label{sec:eval-198}

Order 198 is our hardest exact workload and our most direct test against prior work: one D-optimal design was known, found by a search of about 3,174 worker-hours (104\,h on 32 cores)~\cite{r27,r28}. We asked whether three changes would find more: making the exact test use the server's hardware efficiently, choosing work by join volume, and coordinating the search over a network. Each parent is a pair of length-11  compressions; its units (length-33 pairs, split into classes) each receive an exact test that finds every design they contain. Jobs take parents from a fixed numbering of all 1,020,960 parents, so no two jobs repeat work. We first ran on the 64-thread server, then on 384 FABRIC cores at two sites (\S\ref{sec:fabric}). Every design was verified independently, and before every run, on every machine, a known-answer test re-found the one previously known design~\cite{r27} through the full pipeline, so that a bug could not silently miss designs.

\begin{table}[t]
  \caption{The exact test before and after engineering, on the 64-thread server at full load, same seeds and units.}
  \label{tab:checker}
  \small
  \begin{tabular}{@{}lrr@{}}
    \toprule
     & baseline & engineered \\
    \midrule
    exact-test CPU, same 26 units & 42,276\,s & 11,051\,s (3.83$\times$) \\
    peak memory per process & 2.9\,GB & 0.65\,GB \\
    throughput, 64 vs.\ 32 jobs & 0.95$\times$ & 1.20--1.25$\times$ \\
    \bottomrule
  \end{tabular}
\end{table}

\para{Engineering the exact test.} The careful filter and compact storage of \S\ref{sec:design} made the exact test $3.83\times$ faster on identical work, with output identical to the baseline's, and cut memory $4.5\times$, so
every hardware thread can run a job (Table~\ref{tab:checker}). Because the engineered test gains 20--25\% from the second hardware thread of each core while the (branch-heavy, memory-hungry) baseline loses 5\%, CPU-hours are not comparable across codes.  We therefore report wall-clock time on a stated number of physical cores. Screens that tried to skip work did not pay (Appendix~\ref{app:negative}).

\para{Choosing work by join volume: coordinated by one number.} A class whose two lists have sizes $N_A$ and $N_B$ offers $N_AN_B$ candidate pairs for $N_A+N_B$ work, and the published search already ranked classes this way as a heuristic~\cite{r27,r28}. We measured that solutions grow in proportion to this volume (slope 0.96--0.97 at $v=45$ and 63), which turns the heuristic into a predictive model. Testing only classes above a volume threshold yields $6.1$--$6.3\times$ more expected designs per CPU-hour than testing all work in order.  The best threshold is nearly the same in every parent, so a job that ranks only its own
parent comes within 3\% of a global ranking. The published search's central pool and driver can therefore be replaced
by one broadcast number, so jobs exchange no data at all.

\begin{table}[t]
  \caption{The order-198 campaign by platform and NES software version. Work is measured in \emph{uniform-equivalent units}: the number of units a search examining work uniformly would need to test to cover the same expected number of designs, allowing versions that prioritize work to be compared with those that do not. A unit is the smallest exactly tested task: one candidate class within one length-33 pair of a parent. CPU\,s/unit denotes CPU seconds per unit. Server rows were run on the 64-thread server; FABRIC rows ran the final server version unchanged across two testbed sites.}
  \label{tab:campaign}
  \small
  \resizebox{\columnwidth}{!}{\begin{tabular}{@{}lrrrr@{}}
    \toprule
    Platform and version & CPU-h & units & s/unit & designs \\
    \midrule
    Server v1 (baseline) & 253 & $1.1\cdot10^6$ & 0.48 & 0 \\
    Server v2--v3 (engineered test) & 2,292 & $4.3\cdot10^7$ & 0.19 & 0 \\
    Server v4 (join-volume selection) & 3,722 & $4.1\cdot10^8$ & 0.033 & 1 \\
    Server v5, v5b (shuffled order) & 5,802 & $8.0\cdot10^8$ & 0.026 & 1 \\
    FABRIC, 32-host site & 5,516 & $1.0\cdot10^9$ & 0.019 & 5 \\
    FABRIC, 16-host site & 12,284 & $2.3\cdot10^9$ & 0.019 & 11 \\
    \midrule
    Total & 29,869 & $4.6\cdot10^9$ & & 18 \\
    \bottomrule
  \end{tabular}}
\end{table}

\para{Eighteen new designs.} The search found 18 new D-optimal designs of order 198 (Table~\ref{tab:campaign}). Together with the previously published design, all 19 attain the determinant bound and are pairwise inequivalent as both sequences and matrices (Appendix~\ref{app:designs}). Two designs were found on the server: the first after 3,644 CPU-hours of selective search, and the second, after introducing a per-job randomized parent order, after 477 thread-hours (11.8 wall-clock hours on 48 threads), compared with 104 hours on 32 cores for the published search. The remaining 16 designs came from the FABRIC network-coordinated search, which examined about \(3.4\times10^9\) units in 17,800 CPU-hours, including a 48-hour period on 256 cores that produced 10 designs.

\para{Same code, different luck.} The server found 2 designs in about \(1.2\times10^9\) units, whereas FABRIC found 16 in \(3.4\times10^9\), roughly \(3\times\) the work but \(8\times\) the designs. Was the server less effective, or simply less lucky (luck here meaning finding a design unusually quickly)? The counts are consistent with chance (one-sided binomial test, \(p\approx0.11\)), and determinism lets us test this directly. Twenty-four server jobs reproduced their logs exactly when rerun from their seeds; three FABRIC jobs that found designs reproduced the same designs after the same number of units when rerun on the server; and the FABRIC source builds a byte-identical binary on both platforms. The two systems therefore executed the same search, making chance the most plausible explanation for the difference in designs per unit. Across all runs, the average yield is about one design per \(2.5\times10^8\) units, with substantial uncertainty, so we report design counts and CPU-hours rather than a nominal cost per design.

\begin{figure*}[t]
  \centering
  \includegraphics[width=0.92\textwidth]{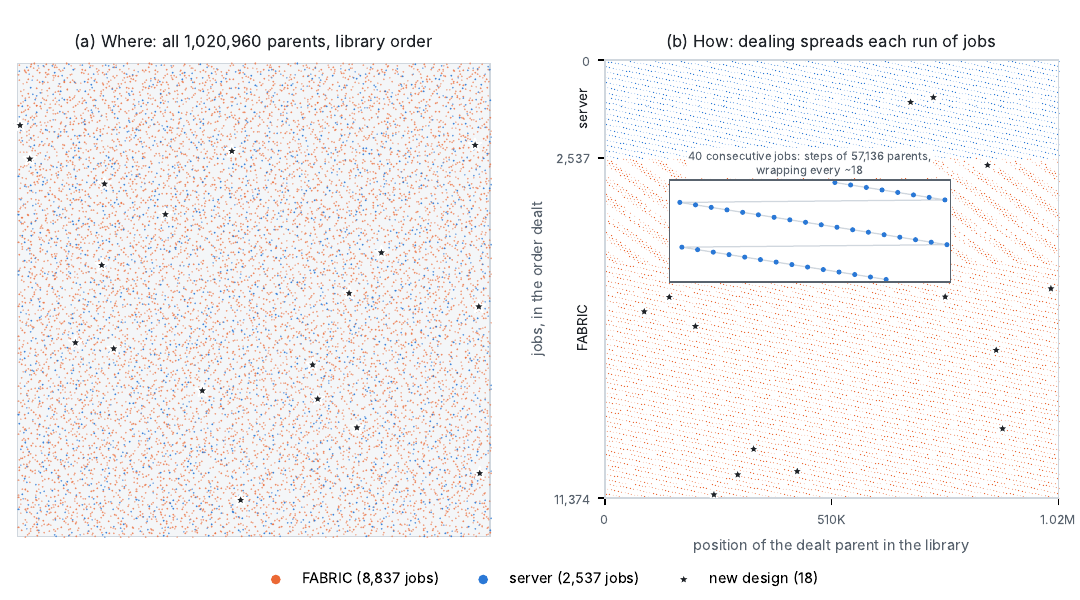}
  \caption{The order-198 search space. (a) All 1,020,960 parents in library order on a Hilbert curve; dots mark the
  11,374 parents searched (each about one fifth; FABRIC orange, server blue); stars are the 18 new designs. (b) How work is dealt: each job's parent is a fixed function of its seed, so consecutive jobs step 57,136 parents through the library and wrap about every 18 jobs (inset). Every run of jobs therefore samples the whole space evenly.}
  \label{fig:coverage}
\end{figure*}

\begin{figure*}[t]
  \centering
  \includegraphics[width=0.92\textwidth]{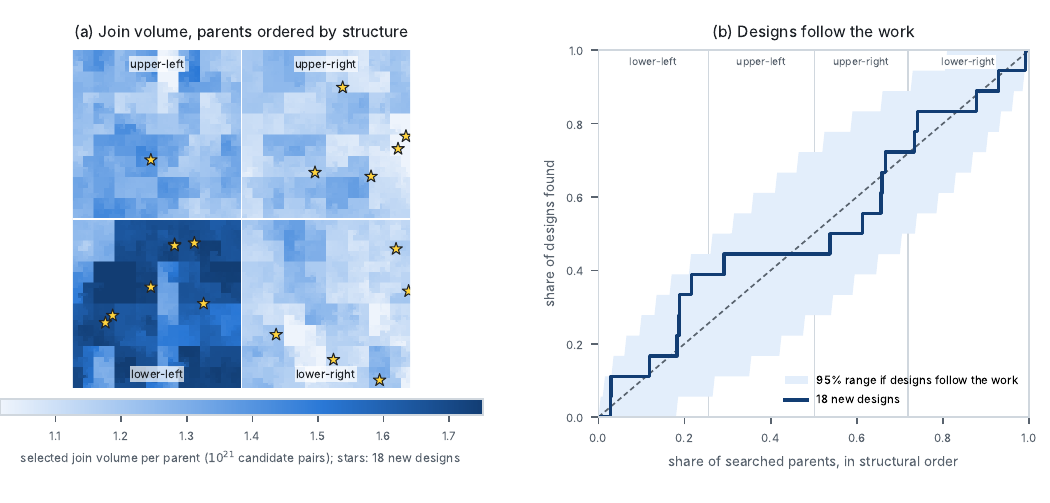}
  \caption{Structure in the order-198 search space. (a) Selected join volume per parent, with parents ordered by their compressed values along a 22-dimensional Hilbert curve~\cite{r80} and smoothed over structural neighbours (an even sample of 4,096 parents); stars are the 18 new designs. One family (lower left) offers about 30\% more selected work per parent. (b) Cumulative designs against cumulative work along the same order; the band is the 95\% range if every job has equal odds. Designs track the work rather than the volume.}
  \label{fig:structure}
\end{figure*}

\para{Where the search went.} Because each job's parent follows from its seed, the campaign's coverage can be mapped
exactly (Figure~\ref{fig:coverage}). The 11,374 parents searched, each about one fifth, are 1.1\% of all parents, and the dealing spreads every run of jobs evenly over the whole space, so both platforms sampled it the same way. Ordering parents by their compressed values along a Hilbert curve~\cite{r80} reveals strong structure in join volume
(Figure~\ref{fig:structure}a): neighboring parents' volumes correlate ($r=0.41$), and one family offers about 30\% more selected work per parent. Yet jobs processed work at the same rate everywhere, and designs appeared in proportion to the work done, not the volume (Figure~\ref{fig:structure}b). Design parents were somewhat richer than typical ($p=0.01$ before correction), a hypothesis we leave open. No region rewarded being searched first, which is why work can be dealt without coordination.

\para{No network on the critical path.} Order 198 sits in the opposite regime to Hadamard. Double compression
partitions the space exactly, so candidates from different units can never match, pooling adds no hits, and 96.5\% of
CPU is one local computation. The network's role is coordination only: a library broadcast, dealt seeds, short result records and a stop signal (\S\ref{sec:fabric}). Exact tests can also be split for scheduling: at D-optimal $v=87$, splitting one parent's test over 8 workers ran $7.1$--$7.6\times$ faster with identical results. Fine exact partitions thus yield independent, compute-bound jobs, while sampled joins at the bottleneck yield pooled, network-amplified rounds.

\subsection{What the search needs from the network}
\label{sec:eval-network}

Our testbed measurements (\S\ref{sec:fabric}) cover CPU nodes and up to 32 hosts. To ask what the network must provide for accelerated nodes and thousands of hosts, we use a model parameterized by measured quantities: the $v=80$ hit
curve ($\lambda=21.5$ at $N=32$\,K, exponent 4), the cost per round ($70\,\mathrm{s}\cdot(N/32\mathrm{K})^{1.9}$ on
one core), $\Nstar$ growing $3.2\times$ per $+8$ in $v$ ($\approx1.5\cdot10^5$ at $v=96$), and 48\,B received per
candidate. A round takes $t_{\mathrm{comp}}(N)/(sP)+t_{\mathrm{net}}$ in Eq.~\ref{eq:ET}, where $s$ is the acceleration of compute and each node receives $V$ bytes over bandwidth $b$. For a non-blocking fabric, a 4:1 oversubscribed packet fabric, and an optical circuit switch (OCS) with 8 circuits per node needing $M=\lceil(P-1)/8\rceil$ matchings, each costing a reconfiguration delay $\delta$,
\begin{equation}
  t_{\mathrm{net}}=
  \begin{cases}
    V/b & \text{non-blocking,}\\
    4V/b & \text{packet, 4:1 oversubscribed,}\\
    \tau + M\delta + V/b & \text{reactive OCS (observes demand for $\tau$),}\\
    (M-1)\delta + V/b & \text{announced OCS (first matching set).}
  \end{cases}
  \label{eq:tnet}
\end{equation}
Announcing demand removes $\tau$ and one reconfiguration per round, so it approaches the ideal fabric only when
$(M-1)\delta$ is small next to compute. On 32 real hosts, an emulation of these schedules matched this accounting
within a few milliseconds (\S\ref{sec:fabric}).

\begin{figure}[t]
  \centering
  \includegraphics[width=\columnwidth]{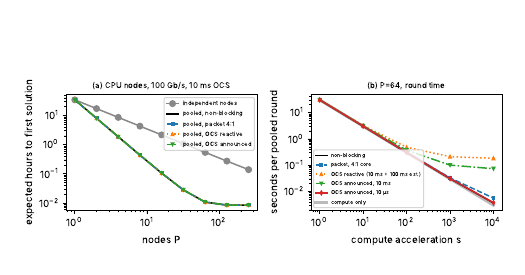}
  \caption{Model (projection from measured parameters), Hadamard $v=96$. (a) Expected time to first solution vs.\ nodes at equal per-node work; independent nodes vs.\ pooled lists on four fabrics (100\,Gb/s, $\delta=10$\,ms,
  $\tau=100$\,ms). (b) Pooled round time at $P=64$ vs.\ compute acceleration.}
  \label{fig:model}
\end{figure}

\begin{table}[t]
  \caption{Model: per-node exchange at $\Nstar$ and compute per pooled round per node at $P=64$ ($v\ge88$
  extrapolated).}
  \label{tab:exchange}
  \small
  \begin{tabular}{@{}rrrrrr@{}}
    \toprule
    $v$ & $\Nstar$ & MB/node & ms at 100G & compute $s{=}1$ & compute $s{=}10^3$ \\
    \midrule
    64 & 1.5\,K & 0.07 & 0.006 & 2.5\,ms & -- \\
    72 & 4.3\,K & 0.20 & 0.016 & 21\,ms & -- \\
    80 & 15\,K & 0.71 & 0.057 & 0.25\,s & 0.25\,ms \\
    88 & 48\,K & 2.3 & 0.18 & 2.5\,s & 2.5\,ms \\
    96 & 154\,K & 7.3 & 0.58 & 25\,s & 25\,ms \\
    \bottomrule
  \end{tabular}
\end{table}

\para{Pooling dominates independence.} At equal per-node work, starting from single-node lists of $\Nstar/8$, pooled
nodes reach a first solution {\bf 8.4$\times$, 16.6$\times$ and 44$\times$ sooner} than independent nodes at $P=8$, 16 and 64, with $\Nstar$ at $v=96$ recalibrated from measured joins to $1.85\cdot10^5$ (the extrapolated $1.5\cdot10^5$ of
Figure~\ref{fig:model}a and Table~\ref{tab:exchange} gives 9.9--50.8$\times$). The gain peaks where the pooled list
reaches $\Nstar$ (here $P\approx50$) and then declines, because extra hits are wasted while independent nodes keep
improving; this is the controller's list-size decision made visible.

\para{On CPU nodes bandwidth barely matters; on accelerated nodes reconfiguration delay does.} At $v=96$ each node
receives 7.3\,MB per round, 0.58\,ms at 100\,Gb/s against about 25\,s of compute on today's CPUs
(Table~\ref{tab:exchange}), so every fabric, even a 4:1 oversubscribed packet core, is within 1\% of ideal
(Figure~\ref{fig:model}a), as compute-bound FABRIC runs confirm (\S\ref{sec:fabric}). At $s=10^3$, compute falls to
25--30\,ms per round and the reconfiguration terms of Eq.~\ref{eq:tnet} dominate (Figure~\ref{fig:model}b): a reactive millisecond-class OCS makes rounds $7\times$ longer than ideal and an announced one $3.3\times$, while the 4:1 packet core costs only 6\% and an announced OCS with 10\,\textmu s switching comes close to ideal. As compute accelerates, reconfiguration delay and acting on announced demand become a fabric's main requirements; bandwidth matters once the exchange is on the critical path, as link-rate measurements on real hosts show (\S\ref{sec:fabric}).

\para{At scale, the network limits problem size.} Using early stopping and flow-level simulation, we project to 8,192 GPU nodes and compare a standard leaf-spine network with three circuit-switched designs: a reactive OCS~\cite{r25,r73}, a fixed-pattern rotor~\cite{r50,r51}, and an announced-demand fabric. Two results stand out. \emph{First, pooled memory sets the largest solvable problem:} independent nodes reach only $v\approx141$--155, while a list-partitioned exchange, in which each node stores $1/P$ of the pooled lists, reaches $v\approx161$--177, so the network acts as a memory bus. \emph{Second, the type of fabric matters most among circuit designs:} at 1,024 nodes, announced circuits are close to ideal and about $5\times$ faster than a reactive OCS. Other differences are smaller. Because expected time grows only
as the $2/3$ power of per-candidate overhead, even an ideal fabric gains little; a leaf-spine with a rack-aware
exchange comes within a few percent of announced circuits; and under background load only reserved announced circuits
are unaffected. Pooling across sites is slower than separate site-local pools, since long-haul costs outweigh the gain and useful pools already fit within one site, and for the order-198 search the network is a negligible share of time to solution.

\begin{table}[t]
  \caption{Model: pooled round time (ms) at $v=96$, $P=64$, $s=10^3$, 100\,Gb/s; ideal fabric 31.1\,ms.}
  \label{tab:sensitivity}
  \small
  \begin{tabular}{@{}lrrrr@{}}
    \toprule
     & \multicolumn{3}{c}{announced, circuits per node} & reactive, 8 circuits \\
    \cmidrule(lr){2-4}\cmidrule(l){5-5}
    $\delta$ & 8 & 32 & 63 & $\tau=1$\,ms / 100\,ms \\
    \midrule
    10\,\textmu s & 31.1 & 31.1 & 31.1 & 32.2 / 131 \\
    100\,\textmu s & 31.8 & 31.2 & 31.1 & 32.9 / 132 \\
    1\,ms & 38.1 & 32.1 & 31.1 & 40.1 / 139 \\
    10\,ms & 101 & 41.1 & 31.1 & 112 / 211 \\
    \bottomrule
  \end{tabular}
\end{table}

\para{Sensitivity.} Since circuit fabrics differ mainly in how fast they reconfigure and how many circuits each node has, we ask how much of each a fabric needs before it stops slowing the search. Varying reconfiguration delay and circuits per node at $s=10^3$ and $P=64$
(Table~\ref{tab:sensitivity}) shows that \emph{with announced demand, delays up to 100\,\textmu s cost at most 2\%}; with a fully connected fabric (63 circuits per node, $M=1$ in Eq.~\ref{eq:tnet}) even 10\,ms switching reaches ideal, since the one matching is set up during computation. A reactive fabric instead pays its observation window every round, and bandwidth and oversubscription matter less than either.

\para{Choosing $N$ online.} Simulated with 64 pooled nodes on the measured $v=80$ hit curve, the controller
(\S\ref{sec:design}), starting $16\times$ below $\Nstar$ and growing $N$ after each empty round, \emph{reaches a first solution in 0.66\,s against 96\,s for a fixed list}, close to the best size chosen in hindsight. Because the list size also fixes the next exchange's volume, the decision that maximizes the odds of success also tells the fabric what to provision.

\para{Small exact jobs.} \para{Small exact jobs.} The rest of this section models the bulk exchange of key lists; the other traffic the search produces (\S\ref{sec:design}) is many small exact tests, each with an input of a few bytes. These need only low latency and fan-out, which any packet network provides. The second traffic class is light: an order-198 unit's input is a few indices, and a D-optimal
slice's input is two short compressed sequences. Its network requirement is latency and fan-out, which a packet fabric
provides.

%% file: sections/07-fabric.tex
%% ---------------------------------------------------------
\section{Network Testbed Measurements}
\label{sec:fabric}

The results so far come from one server and an emulated exchange. To test whether the pooled join behaves the same
over a real network, and what a real network adds, we ran tests on the FABRIC~\cite{r75} testbed.

\para{Testbed.} We configured two FABRIC slices (sets of reserved resources): 32 virtual hosts with 4 cores and
16\,GB each (AMD EPYC 7763) at one site, and 16 hosts with 16 cores and 64\,GB each (AMD EPYC 7542) at a second site,
each slice on a private layer-2 network with a configurable link rate. The search code is byte-identical to our single
server's (same source hash, same coarse library), and seeds reproduce: a job re-run from its seed on the server
re-finds the same order-198 design through the same parent and units. Single-thread speed is similar (the same
exact-test step takes 4.4\,s idle and 5.1\,s loaded on FABRIC against 3.6\,s and 7.9\,s on the server), so differences between platforms are speed, not search behavior.

\begin{table}[t]
  \caption{Pooled vs.\ independent hosts on FABRIC, Hadamard $v=96$, equal total work (pooled list $N_0/\sqrt P$ per
  host). Hits per round and median time to first solution (s). ```Tilted'': each list keeps only the spectrally flattest 30--50\% of candidates, which are more likely to belong to a solution (\S\ref{sec:eval-pooling}).}
  \label{tab:e1}
  \small
  \begin{tabular}{@{}rrrrrr@{}}
    \toprule
    & & \multicolumn{2}{c}{hits per round} & \multicolumn{2}{c}{pooled time (s)} \\
    \cmidrule(lr){3-4}\cmidrule(l){5-6}
    $P$ & pooled $N$ & tilted & untilted & tilted & untilted \\
    \midrule
    4 & 49,152 & 8.0 & 0.3 & 99 & 99 \\
    8 & 34,756 & 14.8 & 0.8 & 109 & 97 \\
    16 & 24,576 & 31.3 & 1.2 & 114 & 107 \\
    32 & 17,378 & 59.9 & 2.5 & 134 & 129 \\
    \bottomrule
  \end{tabular}
\end{table}

\para{Pooled hits scale as predicted over a real network (E1).} We first repeated the pooling emulation of
\S\ref{sec:eval-pooling} on FABRIC, at Hadamard $v=96$ with total work held fixed. \emph{Hits per pooled round grow as
$P^{0.97\text{--}1.02}$ from 4 to 32 hosts} (Table~\ref{tab:e1}), matching the emulation's $P^{0.99}$. \emph{But at this size the extra hits do not shorten the search:} a single host's list is already near $\Nstar$, so pooled hosts took 99--134\,s to a first solution, while independent hosts with full-size lists were often faster. This is the limit of pooling described in \S\ref{sec:eval-pooling}: pooling shortens the search only when single nodes are far below $\Nstar$. 

\begin{table}[t]
  \caption{Splitting one parent's final join across hosts (E14), Hadamard $v=104$, 4 ranks per host. Median time to
  first solution over the 6 parents run at every size; speedup over 1 host: geometric mean of per-parent ratios, with
  bootstrap 95\% CI; exchange share of time.}
  \label{tab:e14}
  \small
  \begin{tabular}{@{}rrrrr@{}}
    \toprule
    hosts & median & speedup (95\% CI) & efficiency & exchange \\
    \midrule
    1 & 72.7\,s & -- & -- & -- \\
    8 & 12.8\,s & 4.6$\times$ (2.6--8.1) & 0.58 & 6\% \\
    32 & 10.6\,s & 5.9$\times$ (3.5--9.1) & 0.18 & 26\% \\
    \bottomrule
  \end{tabular}
\end{table}

\para{Compute-bound rounds confirm the model's CPU prediction.} With the exchange off the critical path, the network had no measurable effect: at $v=96$ and 16 hosts, link rates from 0.05 to 25\,Gb/s, three exchange schedules and two record formats all gave the same time to solution, because the exchange is a few short bursts per round, far below link rate. This confirms on real hosts the model's prediction that every fabric is within 1\% of ideal on CPU nodes (\S\ref{sec:eval-network}), and it is the control for Table~\ref{tab:e16}: the same link-rate sweep, run once the
exchange is on the critical path, changes time to solution by up to $2.2\times$.

\para{Splitting the final join across hosts.} At $v=104$ the final join is the whole cost and a single host's list is
far below $\Nstar$ ($\lambda(N_0)\ll1$), the regime in which the model predicts pooling shortens time to solution. We split one parent's join across hosts (\S\ref{sec:impl}) and compared it with one host and with the same hosts
searching independently (Table~\ref{tab:e14}). \emph{Pooling across 32 hosts reached a first solution $5.9\times$ faster than one host} (median 72.7\,s to 10.6\,s), and the exchange, about 214\,MB per node per round, took a quarter of the time: the first point at which the network is visibly on the critical path. \emph{Pooled hosts were also $1.6\times$ faster than the same hosts searching independently} (median 14.7\,s against 23.2\,s for 8-host runs across 11 parents; one-sided Mann--Whitney $p=0.014$), so the gain comes from pooling, not just from using more hosts.

\para{Coordinating an exact search over the network.} We ran the order-198 search on all 384 cores of both slices,
with one coordinator per site dealing out work and collecting results (\S\ref{sec:impl}). \emph{Coordination cost almost nothing:} sending every node the library of coarse solutions took 0.1--0.2\,s, a stop signal reached every host in a median of 52\,ms, and the only other traffic was a $\sim$100-byte record per finished job. \emph{Over three days this search found 16 of our 18 new designs} (\S\ref{sec:eval-198}).

\begin{table}[t]
  \caption{Link rate when the exchange is on the critical path: split join, $v=104$, parents 1--6, 4 ranks per
  host, each host's egress capped by a token-bucket shaper. Median time to first solution (ratio to 25\,Gb/s), and
  the 32-host exchange time measured and as bytes over the link (856\,MB per host per round).}
  \label{tab:e16}
  \small
  \resizebox{\columnwidth}{!}{\begin{tabular}{@{}rrrrr@{}}
    \toprule
    cap per host & 32 hosts & 8 hosts & exchange, 32 hosts & bytes/link \\
    \midrule
    25\,Gb/s & 10.4\,s & 13.0\,s & 2.54\,s & 0.27\,s \\
    10\,Gb/s & 10.2\,s (0.99) & 12.7\,s (0.98) & 2.54\,s & 0.68\,s \\
    5\,Gb/s & 9.9\,s (0.95) & 13.0\,s (1.00) & 1.99\,s & 1.37\,s \\
    2\,Gb/s & 11.6\,s (1.11) & 15.6\,s (1.20) & 3.81\,s & 3.42\,s \\
    1\,Gb/s & 15.5\,s (1.49) & 18.5\,s (1.42) & 7.45\,s & 6.85\,s \\
    0.5\,Gb/s & 22.6\,s (2.17) & 25.1\,s (1.93) & 14.74\,s & 13.70\,s \\
    \bottomrule
  \end{tabular}}
\end{table}

\begin{figure}[t]
  \centering
  \includegraphics[width=\columnwidth]{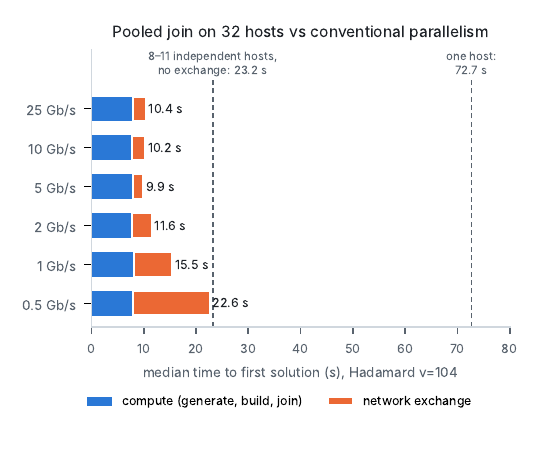}
  \caption{Time to the first solution for the pooled join on 32 FABRIC hosts (Hadamard $v=104$, parents 1--6, medians),
  split into compute and network exchange, at each per-host link rate (E16). Dashed lines: the same algorithm on one host
  (72.7\,s) and 8--11 hosts searching independently with no exchange (23.2\,s), measured on the same testbed.}
  \label{fig:linkrate}
\end{figure}

\para{Once the exchange is on the critical path, the network sets time to solution.} We repeated the 32- and 8-host split joins with each host's egress capped from 25 down to 0.5\,Gb/s (Table~\ref{tab:e16}). \emph{As the cap falls, time to a first solution grows by up to $2.2\times$}, with compute fixed near 8\,s and the exchange growing from a quarter to two thirds of the round (Figure~\ref{fig:linkrate}). Above 5\,Gb/s the exchange is limited by how it is scheduled rather than by link capacity, which leaves room for a better schedule. \emph{The network also decides whether pooling helps:} at 2--25\,Gb/s the pooled join is $2.0$--$2.3\times$ faster than the same hosts searching independently, but at 0.5\,Gb/s that advantage disappears. {\em Together, pooling and splitting the join make the search $3.2\times$ faster than a single host at 0.5\,Gb/s and about $7\times$ faster at 5--25\,Gb/s (Figure~\ref{fig:linkrate}): the same algorithm on the same hosts gains more than twice as much from a fast network as from a slow one.}

\para{Schedule accounting on real hosts.} To test the model's accounting of circuit schedules (Eq.~\ref{eq:tnet}), we emulated them on all 32 hosts: each round, hosts compute for a set time and then exchange real data over TCP, either all at once or through four configurations of 8 partners each, with circuit setup emulated by a synchronized pause of $\delta$. This is an emulation on a packet network, not an optical fabric. \emph{Every schedule added what the model predicts, within a few milliseconds}, for announced, rotor and reactive schedules alike. \emph{Announced demand hid exactly one reconfiguration:} with $\delta=10$\,ms it added three delays rather than four, because the first configuration was set up during compute. At 25\,ms of compute, the accelerator-like regime where these differences matter most, the emulator's own variability masked them; measuring that regime needs a native implementation.

%% file: sections/08-discussion.tex
%% ---------------------------------------------------------------- 8 Discussion
%% Updated 19 Nov: claims reflect FABRIC results and the eighteen designs; "planned experiments" -> completed/remaining;

\section{Discussion and New Directions}
\label{sec:discussion}

\para{What we can and cannot claim.} \emph{First, restructuring search to exploit the network yields significant speedups.} Pooling increases hits per unit of work by roughly \(P\), and on FABRIC the pooled join reached a first solution \(5.9\times\) faster than a single host and \(1.6\times\) faster than the same hosts searching independently; the network itself determined the gain, from \(3.2\times\) at 0.5\,Gb/s to about \(7\times\) at 5--25\,Gb/s. \emph{Second, the search discovered eighteen new order-198 D-optimal designs}, showing that the method can reach this order repeatedly rather than as an isolated success. \emph{Third, it may hold the time record for this order:} our second design was found after 11.8 hours of wall-clock time on 48 threads, compared with 104 hours on 32 cores for the published search~\cite{r27}, to our knowledge the fastest previously reported result. Several limitations remain.  We claim no new orders: at the open D-optimal order 270, a pilot study estimates about \(10^{10}\) thread-hours to the first design (Appendix~\ref{app:negative}), while the smallest open Hadamard orders lie well beyond the measured growth curve. Pooling itself has two fundamental limits. \emph{Saturation:} once a pooled round succeeds with high probability, additional nodes increase the number of solutions found but not the time to the first one (Figure~\ref{fig:model}a). \emph{Scarce parents:} when extendable parents are rare (about 2\% for D-optimal \(v=69\)--87), the critical path is finding one, and many independent exact tests distributed across nodes hedge this uncertainty better than a single pooled join.

\para{Deploying pooled joins on large GPU pools.} Two issues shape a large deployment. \emph{Stragglers:} a pooled
round is a bulk-synchronous step, so the slowest node sets its pace, and on shared GPU nodes at $P\ge1{,}024$ this makes
time to first solution 1.2--2.7$\times$ longer. Asynchronous rounds, which close when nearly all data has arrived and
carry late data into the next round, recover almost all of this in our model while still joining every candidate
exactly once. \emph{Pool size:} only nodes working on the same parent can combine candidates (joins across parents
found no hits), so a large deployment is many per-parent pools rather than one, and each parent's completion space
bounds how large its pool can usefully grow. A parent can announce that bound in a few bytes, letting a scheduler size
and place pools. Pooling also aggregates memory: lists that need 1.4\,GB on one node need a few tens of MB per node
across 64 nodes. Because rounds are memoryless, failures are cheap: a failed node's buckets can be reassigned and the
round rerun with no loss of correctness, and keeping parents across rounds avoids repeating the coarse search.
Asynchronous, parent-aware and fault-tolerant rounds are the next steps toward GPU clusters of tens of thousands of
nodes, where the open D-optimal order 270 (about $10^{10}$ thread-hours by our pilot) would take weeks to months, a
projection that brings new orders within reach.

\para{Experiments completed and remaining.} Our FABRIC experiments validate the model of \S\ref{sec:model}: hits scale as \(N^{\alpha}\), work as \(N^{\beta}\), and pooling therefore yields an approximately linear gain in \(P\). Pooled hits increase in proportion to host count, and pooling reduces time to solution only when individual hosts operate far below \(\Nstar\): it provided no benefit at \(v=96\) but a \(5.9\times\) speedup at \(v=104\). When computation dominates, link rate is irrelevant; once communication reaches the critical path, it changes time to solution by up to \(2.2\times\), and above 5\,Gb/s performance is limited by exchange scheduling rather than bandwidth. On real hosts, emulated circuit schedules closely matched the model's accounting of reconfiguration delay and announced demand (Eq.~\ref{eq:tnet}, \S\ref{sec:fabric}). Four questions remain. Can announced demand hide reconfiguration delay at accelerator speeds, where it matters most? Can a hit-rate-driven controller adapt list sizes and announce demand on real hosts while operating far below \(\Nstar\)? Does pooling's advantage over independent hosts continue to grow at \(v=112\), where hosts are farther from \(\Nstar\)? And do asynchronous rounds and list-partitioned exchange remain effective under injected delay and background traffic? Porting the join to GPUs makes these questions decisive: with computation accelerated by \(10^2\)–\(10^3\), the network moves onto the critical path.

\para{Building the controller.} The controller of \S\ref{sec:design} makes two decisions that can run in different places. The solver, the search program on each node, chooses the list size from each round's hits. Before each exchange it would announce participants, volume and start time to a software-defined networking controller or an optical switch's management interface, which turns them into circuit matchings or paced, reserved paths. Since the exchange repeats every round with a known shape, the schedule can be computed once per pool and re-timed each round.  On packet networks, pacing alone may recover much of the scheduling loss measured above 5\,Gb/s. Linking announcements to cluster schedulers would let one interface size, place and reserve each pool.

\para{Toward open orders.} The open Hadamard orders remain far beyond today's frontier. Order 2396, for example, requires length \(v=598\), versus \(v=120\) today; extrapolating the measured growth curve, a generic search would cost roughly \(10^{51}\) times more. Pooling contributes an approximately linear gain in \(P\), enough to recover only tens of units of length. Reaching open orders will therefore require algebraic structure, not scale alone. \emph{The network then becomes memory.} Algebra can restrict a prime length such as 599 to sequences built from symmetry orbits, and folds continue to partition the resulting search space, but even these restricted searches leave lists of \(10^{10}\)--\(10^{11}\) keys per sequence, far beyond the capacity of a single node. Such lists are joinable only if memory is pooled across nodes, precisely the regime in which our model predicts that list-partitioned exchange, and the fabric supporting it, determine how far a search can reach (\S\ref{sec:eval-network}). \emph{Calibrate, then commit.} The real uncertainty is whether a chosen symmetry class contains any solutions at all. At D-optimal \(v=99\), every symmetry subgroup we tested was empty (Appendix~\ref{app:negative}). Because hits are observable each round and waiting times are memoryless, the hit curve can be calibrated on smaller analogues with known solutions before committing cluster-scale resources. The order-198 campaign further shows that, once calibrated, an exact search can run for months across software versions, platforms, and sites while remaining reproducible, and that even a search that finds nothing still produces a quantitative bound.

\para{A research agenda for Search-Defined Networking.} Three directions follow from this work. \emph{Communication-amplified computation beyond Hadamard.} Any search whose success probability grows faster with pooled state than its work, and whose exchanged state is immediately usable, belongs to this class. Candidate domains include subset sum, code- and lattice-based cryptanalysis, and other combinatorial design problems (\S\ref{sec:generality}, Table~\ref{tab:generality}). Comparing against fixed baselines will reveal how broad the class is. \emph{Announced-demand fabrics.} Search is not unique in knowing its next exchange in advance: iterative solvers and synchronous training workloads do as well. An interface for announcing demand would let circuit fabrics configure during computation. In multi-tenant settings, list size itself becomes a network resource, trading communication against success probability. \emph{AI that shapes the search.} Our results identify where AI can help. Waiting times are memoryless and parent features weakly predictive, so forecasting successful runs offers little value. Instead, AI should increase success probability per unit of work by designing compression towers, keys, join orders, and solution-biased samplers. Because hits are observable every round, such proposals can be evaluated on small analogues with known solutions before deployment at scale, creating a rapid, measurable loop of design, search, and evaluation.

%% file: sections/09-related.tex
%% ---------------------------------------------------------------- 9 Related Work
\section{Related Work}
\label{sec:related}

\para{Demand-inferred networks.} Hedera reschedules large flows~\cite{r3}, and Helios and c-Through add optical
circuits for heavy demand estimated from traffic counters~\cite{r25,r73}. Related systems reduce control overhead, exploit short-term predictability, make the optical core flexible, or allocate WAN bandwidth centrally
\cite{r21,r10,r14,r33,r36}. All of them learn demand after it appears. In NES the application announces demand before
the exchange starts, and its size is itself a control decision.

\para{Fast and demand-oblivious reconfiguration.} Mordia brought microsecond circuit switching to the data
center~\cite{r59}, and RotorNet and Opera rotate through matchings independently of demand~\cite{r50,r51}. Other
designs use free-space optics or rotor-style switching, and optical circuit switches are deployed in production
\cite{r29,r30,r4,r8,r38,r47,r60,r70}. Our model shows that this workload needs fast reconfiguration but not
obliviousness, since announced demand lets even a slower OCS hide part of its delay.

\para{Application-exposed structure.} The coflow abstraction lets an application expose the collective structure of its transfers~\cite{r18}, and Varys and Aalo schedule coflows with and without prior knowledge~\cite{r17,r19}. Related work exposes endpoints, collective transfers or routing to the application \cite{r16,r15,r2}. These systems expose transfers the application has already decided on. In NES the volume is chosen jointly with the network, because it changes the probability of success.

\para{Workload--network co-design for ML.} TopoOpt co-optimizes topology and parallelization~\cite{r74}, and SiP-ML evaluates silicon-photonic fabrics for training~\cite{r41}. Other work schedules jobs to reduce interference or describes production training networks \cite{r13,r62,r26,r61}. Training communication is overhead with a fixed volume per step. Ours is a source of gain whose volume is a tunable parameter.

\para{AI-assisted mathematics.} Program search with language models has produced new extremal
constructions~\cite{r64}, and the order-668 Hadamard matrix was found with an AI assistant~\cite{r39}. These systems
change how constructions are proposed. NES addresses how the resulting searches are partitioned and what the network
does for them, and the two are complementary.

\para{Collision search and generic joins.} Parallel collision search pools lists across machines, with memoryless waits set by list size~\cite{r71}, and Wagner's algorithm is the template for joins over huge lists~\cite{r72}. Our pooled join applies the same principle to sums of correlation vectors and adds that each exchange can be announced to the network. Knapsack, decoding and sieving algorithms share the join structure \cite{r76,r77,r78,r79}, and their distributed versions are natural next workloads.

\para{Parallel combinatorial search and constructions.} Cube-and-conquer settled the Pythagorean triples and Schur
number five~\cite{r31,r32}, and the published order-198 design was found by exact compression~\cite{r27}, the closest construction to ours. Other parallel searches share clauses, distribute hashing, prune with pattern databases or
enumerate isomorph-free objects \cite{r9,r66,r42,r63,r20,r44,r49}. Among single-node heuristics for Hadamard matrices, Kline's log-determinant relaxation found four-circulant matrices up to length 29 within a day~\cite{r43}, while our single-core tower solves the bordered four-circulant problem at length 96. Different structures and hardware make this an indication of scale, not a controlled comparison. Projection methods and early restarts \cite{r7,r23,r34,r40} did not transfer to the tower (Appendix~\ref{app:negative}).

%% file: sections/10-conclusion.tex
%% ---------------------------------------------------------------- 10 Conclusion
%% Updated 19 Nov: FABRIC results, eighteen designs, generality.
\section{Conclusion}
\label{sec:conclusion}

Data-center network research has largely asked how to serve an application’s traffic efficiently, treating that traffic as a cost to be minimized.  Large-scale mathematical search, increasingly driven by AI, has been treated as the opposite case, a collection of independent jobs that ask almost nothing of the network. This paper shows that such searches can be restructured so that communication raises
their probability of success, and calls this \emph{network-enhanced search}. The restructuring applies to searches with an exact reduction to smaller problems, additive keys and sampled joins. Built from compression towers and sampled exact joins, such a search finds about $P$ times more solutions per unit of work when $P$ nodes pool their lists, and it knows the size and timing of every exchange before it happens. On the FABRIC testbed, we found that splitting one join across 32 hosts reached a first solution $5.9\times$ faster than one host and $1.6\times$
faster than the same hosts searching independently, and the network itself decided the gain, from $3.2\times$ on slow
links to about $7\times$ on fast ones. The same search, coordinated over the network, found sixteen of eighteen new
D-optimal designs of order 198, and our measurement-parameterized model shows that at scale pooled memory sets the largest solvable problem, while on accelerated nodes reconfiguration delay and announced demand matter more than
bandwidth. This work opens several new directions, including applying network-enhanced search to cryptanalysis and other combinatorial problems, fabrics that act on announced demand, and AI systems that propose the structure of a search and learn where its solutions live. Combined with algebraic advances that shrink the search space, these directions could bring open orders within reach.

%% file: sections/11-acknowledgments.tex
%% ---------------------------------------------------------------- Acknowledgments
\ifsigcomm\else
\begin{acks}
Experiments, code and drafting were carried out with Claude (Anthropic). The order-668 sequences come from the notebook
released with~\cite{r39}. We thank Stelios Georgiou and Stella Stylianou for details of their order-198 computation.
\end{acks}
\fi

%% file: sections/A-designs.tex
%% ---------------------------------------------------------------- Appendix A: designs
\section{The New Order-198 Designs}
\label{app:designs}

This appendix lists the eighteen new D-optimal designs of order 198, records where each was found so that any of them
can be reproduced from its seed, and gives the evidence that they are valid and pairwise inequivalent.

\para{The construction.} With $v=99$ and two $\pm1$ sequences $A$ and $B$ of length 99, the matrix
$M=\left(\begin{smallmatrix}C_A & C_B\\ -C_B^{\top} & C_A^{\top}\end{smallmatrix}\right)$, built from the circulant
matrices $C_A$ and $C_B$, is D-optimal of order 198 when the agreements of $A$ and $B$ (\S\ref{sec:background}) sum to
2 at every nonzero shift, $P_A(s)+P_B(s)=2$ for $s\ne0$, where $P$ is the periodic autocorrelation. The row sums of
$A$ and $B$ must then be 13 and 15, the only choice for $4v-2=394$.

\para{Where each design was found.} Every design came from an ordinary dealt job, so each is identified by the seed
from which its job's parent and work order follow (\S\ref{sec:eval-198}). Designs 1 and 2 were found on the 64-thread
server on 11 and 12 November 2026. Design 1 came from seed 20973 (parent 739,593, length-33 pair 4,246 of 30,741) and
design 2 from seed 30121 (parent 688,201, pair 2,265 of 33,250 in the job's shuffled order). Designs 3--18 were found
on FABRIC by the network-coordinated search (\S\ref{sec:fabric}) between 16 and 19 November. Design 3 came from seed
54562 on the 32-host site and design 4 from seed 59626 on the 16-host site, after only 1,097 units of its first parent.
Designs 5--14 came from the 16-host site (seeds 58123, 58130, 58540, 58953, 61842, 63049, 63136, 57877, 62404 and
63689) and designs 15--18 from the 32-host site (seeds 52353, 53330, 56507 and 52057). Designs 5--18 are numbered by
site and collection round, not by time of discovery. Below, $+$ denotes $+1$, and each sequence is written in three
rows of 33.

\para{Design 1.}
{\footnotesize\ttfamily
\begin{flushleft}
A = -{}+++-{}+-{}++++-{}-{}+++-{}+-{}++-{}+-{}-{}-{}-{}-{}-{}+++-{}\\
\hspace*{2.6em}-{}+-{}-{}-{}-{}-{}+++++-{}-{}+++-{}+++-{}-{}++-{}+-{}-{}-{}+-{}-{}\\
\hspace*{2.6em}-{}++++++-{}+-{}+-{}+-{}-{}-{}++++-{}++++-{}++-{}++-{}+\\[2pt]
B = -{}+-{}-{}+++++-{}+-{}+++-{}-{}++-{}++-{}++-{}-{}+++++-{}\\
\hspace*{2.6em}-{}-{}-{}-{}+-{}++-{}-{}-{}-{}-{}-{}+-{}+-{}-{}++++++-{}++++-{}-{}+\\
\hspace*{2.6em}-{}+-{}-{}++-{}-{}++++-{}++-{}+-{}++-{}-{}++-{}+++-{}++-{}+
\end{flushleft}}

\para{Design 2.}
{\footnotesize\ttfamily
\begin{flushleft}
A = -{}++++++-{}+-{}-{}-{}-{}++-{}+-{}-{}-{}+-{}-{}+-{}-{}++++-{}++\\
\hspace*{2.6em}-{}-{}-{}-{}+++-{}-{}++-{}-{}-{}+++-{}+-{}-{}-{}+-{}+++-{}-{}++-{}+\\
\hspace*{2.6em}-{}++++++-{}+++-{}+++-{}-{}+-{}+-{}++++-{}-{}+-{}-{}+++\\[2pt]
B = -{}++-{}++-{}++-{}+-{}+++++-{}-{}-{}-{}+-{}-{}-{}++-{}++-{}+-{}\\
\hspace*{2.6em}-{}+++-{}+-{}-{}-{}+++-{}-{}++++-{}+-{}+-{}++-{}+++++++\\
\hspace*{2.6em}++++-{}+-{}+-{}-{}-{}-{}-{}+++-{}-{}+-{}+++-{}-{}+-{}-{}-{}+-{}++
\end{flushleft}}

\para{Design 3.}
{\footnotesize\ttfamily
\begin{flushleft}
A = ++-{}-{}+-{}+-{}++-{}+++-{}-{}++-{}-{}-{}-{}-{}+-{}++-{}-{}-{}+++\\
\hspace*{2.6em}-{}++-{}+++++-{}-{}-{}+++++-{}++-{}-{}-{}-{}+-{}+-{}++-{}-{}+\\
\hspace*{2.6em}-{}+++-{}++-{}-{}+-{}-{}+-{}+-{}+-{}-{}-{}++-{}+-{}++++++++\\[2pt]
B = +-{}+++++++-{}-{}-{}++++-{}++-{}-{}+++++++-{}+-{}++\\
\hspace*{2.6em}-{}-{}-{}-{}++-{}+++-{}+-{}-{}-{}-{}+-{}+++-{}-{}+++-{}+-{}-{}-{}+-{}\\
\hspace*{2.6em}-{}-{}+-{}+-{}-{}++-{}+-{}-{}++++-{}+++-{}-{}+-{}++-{}+-{}-{}++
\end{flushleft}}

\para{Design 4.}
{\footnotesize\ttfamily
\begin{flushleft}
A = -{}++-{}+-{}+-{}+-{}-{}-{}-{}+++-{}+-{}-{}-{}+++-{}+-{}+-{}-{}+-{}+\\
\hspace*{2.6em}-{}++-{}++++++++-{}-{}++-{}+++++-{}+++-{}-{}++-{}+-{}\\
\hspace*{2.6em}-{}++++-{}-{}-{}-{}+-{}-{}+-{}+++-{}-{}++-{}-{}+++++-{}-{}-{}-{}+\\[2pt]
B = -{}++-{}+++++-{}++++-{}-{}+-{}-{}-{}+-{}+++-{}+++++++\\
\hspace*{2.6em}-{}-{}+-{}+-{}+-{}++-{}-{}+-{}++-{}-{}-{}-{}+++++++-{}++-{}++\\
\hspace*{2.6em}-{}+++-{}-{}-{}++-{}-{}-{}-{}-{}-{}++-{}-{}+++-{}+-{}+-{}-{}+-{}++-{}
\end{flushleft}}

\para{Design 5.}
{\footnotesize\ttfamily
\begin{flushleft}
A = -{}-{}+-{}++-{}+++-{}-{}-{}-{}-{}+-{}-{}++++-{}++++-{}-{}+++-{}\\
\hspace*{2.6em}+-{}-{}-{}+++-{}+++-{}+-{}-{}-{}+++++-{}-{}++-{}-{}+-{}+-{}+-{}\\
\hspace*{2.6em}++++-{}+-{}-{}++++-{}++-{}+-{}++-{}-{}-{}+-{}++-{}+-{}-{}++\\[2pt]
B = -{}++-{}+++-{}+-{}++++++-{}-{}++++-{}-{}-{}+++++++-{}\\
\hspace*{2.6em}-{}+-{}-{}+++-{}++++-{}+++-{}+-{}-{}-{}-{}+-{}-{}++-{}+++++\\
\hspace*{2.6em}-{}++++-{}++-{}-{}-{}-{}+-{}+-{}-{}+-{}-{}-{}-{}-{}+-{}+-{}+-{}-{}-{}++
\end{flushleft}}

\para{Design 6.}
{\footnotesize\ttfamily
\begin{flushleft}
A = -{}-{}-{}++-{}+-{}-{}-{}-{}-{}-{}+++++-{}-{}++-{}+-{}++-{}+-{}+-{}-{}\\
\hspace*{2.6em}-{}+++-{}-{}+++-{}++-{}++-{}++-{}++-{}-{}+-{}-{}-{}+-{}+++-{}\\
\hspace*{2.6em}++++-{}+++++-{}+++-{}+-{}+++++-{}+-{}+-{}-{}+-{}-{}-{}+\\[2pt]
B = -{}-{}+++++++-{}-{}++-{}++++-{}-{}+-{}++-{}+-{}-{}+-{}+++\\
\hspace*{2.6em}-{}-{}+-{}+++++-{}+-{}+++-{}-{}-{}++-{}-{}-{}-{}+-{}+++-{}-{}+-{}\\
\hspace*{2.6em}-{}+++-{}+++-{}-{}-{}-{}-{}++-{}++++++-{}+-{}+-{}-{}-{}-{}+++
\end{flushleft}}

\para{Design 7.}
{\footnotesize\ttfamily
\begin{flushleft}
A = -{}-{}+++++-{}+-{}-{}-{}++++-{}+-{}++-{}-{}-{}++-{}-{}+-{}+++\\
\hspace*{2.6em}-{}++-{}-{}+-{}-{}-{}+-{}+-{}-{}-{}-{}+-{}-{}+-{}+++-{}+++-{}+-{}-{}+\\
\hspace*{2.6em}-{}+++++-{}-{}-{}+-{}-{}-{}+++++++-{}++++-{}-{}+-{}++++\\[2pt]
B = -{}-{}-{}-{}+++++-{}-{}-{}-{}+++-{}-{}+-{}++-{}++++-{}+++-{}+\\
\hspace*{2.6em}+-{}-{}++-{}-{}++-{}+-{}+-{}+-{}+-{}-{}++-{}-{}+-{}+-{}-{}++-{}++\\
\hspace*{2.6em}-{}+-{}-{}+-{}+-{}++-{}-{}-{}+++++++++++-{}++++-{}-{}-{}+
\end{flushleft}}

\para{Design 8.}
{\footnotesize\ttfamily
\begin{flushleft}
A = -{}-{}+++-{}++++-{}-{}++++-{}++-{}++++-{}+++-{}-{}-{}-{}+\\
\hspace*{2.6em}+-{}-{}-{}+-{}+++-{}++-{}-{}+-{}+-{}++-{}-{}-{}-{}++++++-{}++\\
\hspace*{2.6em}-{}-{}++-{}+-{}+++-{}-{}-{}-{}+-{}+-{}-{}+-{}-{}-{}+++-{}+-{}+-{}++\\[2pt]
B = -{}++-{}+-{}-{}-{}-{}-{}+-{}-{}-{}++++-{}++-{}-{}-{}+-{}++++-{}+-{}\\
\hspace*{2.6em}-{}-{}-{}+-{}-{}++-{}++-{}++++++-{}-{}-{}+++-{}++-{}+++++\\
\hspace*{2.6em}-{}+++-{}+-{}-{}-{}++-{}+-{}+-{}+++++-{}+++++-{}-{}+-{}-{}+
\end{flushleft}}

\para{Design 9.}
{\footnotesize\ttfamily
\begin{flushleft}
A = -{}-{}-{}+++-{}+++-{}+-{}+++++-{}+-{}-{}-{}++++-{}-{}+++-{}\\
\hspace*{2.6em}-{}-{}++++++-{}+-{}+-{}-{}++-{}-{}+-{}-{}+-{}-{}-{}-{}-{}+++-{}++\\
\hspace*{2.6em}-{}++-{}+-{}++-{}++++++-{}-{}-{}+-{}+-{}-{}+-{}+-{}+++-{}-{}+\\[2pt]
B = -{}++-{}-{}++-{}+-{}++-{}+++-{}-{}+++++++++-{}++++-{}\\
\hspace*{2.6em}-{}+-{}-{}+++++-{}+-{}+++-{}+-{}+-{}-{}-{}-{}++-{}-{}+-{}-{}-{}-{}+\\
\hspace*{2.6em}+++-{}++++-{}-{}-{}-{}-{}++-{}-{}+-{}+++-{}++-{}+-{}+-{}-{}+-{}
\end{flushleft}}

\para{Design 10.}
{\footnotesize\ttfamily
\begin{flushleft}
A = -{}-{}++-{}+-{}++-{}+-{}+++-{}++++-{}-{}-{}-{}+++-{}-{}-{}++-{}\\
\hspace*{2.6em}+++-{}-{}+-{}++-{}-{}-{}-{}+-{}++-{}++++-{}+-{}+-{}-{}-{}+-{}-{}-{}\\
\hspace*{2.6em}+-{}+-{}-{}++-{}++-{}++-{}++++-{}+++-{}-{}-{}+-{}++++++\\[2pt]
B = +-{}+++++-{}-{}-{}++++-{}-{}-{}+-{}-{}+-{}-{}+++-{}++++++\\
\hspace*{2.6em}-{}+++++-{}+-{}-{}+-{}-{}-{}++-{}-{}++-{}-{}-{}-{}-{}+-{}+++-{}++\\
\hspace*{2.6em}-{}-{}-{}-{}-{}+-{}++-{}+++-{}+-{}++-{}+++-{}-{}++++++-{}+-{}
\end{flushleft}}

\para{Design 11.}
{\footnotesize\ttfamily
\begin{flushleft}
A = -{}-{}+++++-{}+-{}+-{}-{}+++-{}++-{}+-{}-{}+-{}+++-{}++-{}+\\
\hspace*{2.6em}-{}-{}-{}+-{}++-{}-{}+++-{}++-{}+-{}+++-{}-{}+++-{}+++-{}++\\
\hspace*{2.6em}+++-{}-{}-{}+-{}+-{}+++++-{}+-{}-{}+-{}-{}-{}-{}-{}-{}-{}++-{}++-{}\\[2pt]
B = -{}-{}-{}+-{}-{}-{}++-{}++-{}-{}++++-{}+++-{}-{}+++++-{}-{}-{}-{}\\
\hspace*{2.6em}-{}++-{}+-{}-{}++++-{}-{}+++++++++-{}+-{}+-{}+-{}+-{}++\\
\hspace*{2.6em}-{}++++++-{}-{}-{}++-{}-{}+-{}++-{}+-{}++-{}-{}-{}++-{}++-{}-{}
\end{flushleft}}

\para{Design 12.}
{\footnotesize\ttfamily
\begin{flushleft}
A = -{}+++++-{}++-{}-{}+-{}+-{}++-{}-{}-{}++-{}-{}-{}+-{}-{}-{}-{}++-{}\\
\hspace*{2.6em}-{}-{}+++++-{}+-{}-{}-{}++-{}-{}+++++-{}-{}+++++++-{}++\\
\hspace*{2.6em}-{}+++-{}+-{}-{}+-{}++++++-{}+-{}+-{}+-{}-{}+-{}-{}+-{}-{}+-{}+\\[2pt]
B = -{}-{}-{}-{}-{}+-{}++++-{}-{}+-{}-{}+++-{}++++-{}++-{}++-{}+-{}\\
\hspace*{2.6em}+++-{}-{}-{}-{}++-{}+-{}+-{}+++-{}-{}+-{}-{}++-{}++++++-{}+\\
\hspace*{2.6em}+-{}++-{}-{}-{}-{}+++-{}+++-{}-{}+++++-{}-{}-{}+++-{}+-{}+-{}
\end{flushleft}}

\para{Design 13.}
{\footnotesize\ttfamily
\begin{flushleft}
A = +-{}-{}+-{}++++-{}+-{}+-{}+-{}+-{}-{}+-{}-{}-{}+++-{}-{}+++-{}-{}\\
\hspace*{2.6em}-{}-{}+-{}+++-{}-{}+-{}-{}+-{}-{}+++++++-{}+++-{}-{}++-{}-{}-{}\\
\hspace*{2.6em}-{}-{}++-{}+-{}-{}+-{}++-{}++-{}+-{}+++-{}++-{}-{}+++++++\\[2pt]
B = -{}+++-{}++-{}-{}+++-{}-{}+-{}++-{}+++-{}+++++-{}-{}+++\\
\hspace*{2.6em}++-{}-{}-{}+-{}+-{}+-{}+-{}++-{}+-{}-{}+-{}+-{}-{}++++++-{}+-{}\\
\hspace*{2.6em}-{}-{}+-{}-{}-{}-{}++-{}+-{}-{}++-{}-{}+++-{}-{}-{}-{}-{}++++++++
\end{flushleft}}

\para{Design 14.}
{\footnotesize\ttfamily
\begin{flushleft}
A = -{}+-{}++-{}+++-{}-{}-{}++-{}+++-{}+++-{}+-{}+-{}+-{}-{}+++\\
\hspace*{2.6em}-{}++-{}+-{}-{}-{}-{}+-{}+-{}-{}++-{}+-{}-{}-{}-{}-{}+++-{}+++++-{}\\
\hspace*{2.6em}-{}+++-{}-{}-{}-{}+-{}-{}+-{}-{}+-{}+++++++++-{}+-{}-{}++++\\[2pt]
B = -{}-{}-{}++++-{}-{}+-{}-{}++-{}+-{}+-{}++-{}-{}++-{}++-{}+-{}+-{}\\
\hspace*{2.6em}-{}+++++++-{}-{}+-{}-{}-{}++-{}-{}-{}+-{}-{}-{}+-{}++++-{}-{}-{}+\\
\hspace*{2.6em}-{}+-{}-{}+++++-{}+++++-{}++-{}+++++-{}-{}-{}+++-{}++
\end{flushleft}}

\para{Design 15.}
{\footnotesize\ttfamily
\begin{flushleft}
A = -{}-{}-{}-{}-{}-{}-{}++-{}++++-{}+++++-{}-{}-{}++-{}+-{}++-{}++\\
\hspace*{2.6em}+-{}-{}++++-{}+-{}++-{}-{}++++-{}-{}+-{}-{}-{}+-{}+++-{}-{}-{}-{}\\
\hspace*{2.6em}-{}++-{}+++-{}-{}++-{}+-{}+-{}++++-{}+-{}++-{}-{}+++-{}++\\[2pt]
B = -{}+-{}-{}-{}-{}+++-{}-{}-{}-{}-{}++-{}+-{}+++-{}+++-{}+-{}-{}-{}+-{}\\
\hspace*{2.6em}++-{}-{}-{}++-{}+-{}+-{}-{}-{}++++-{}++-{}++-{}++-{}-{}+-{}+-{}\\
\hspace*{2.6em}-{}+++++-{}++++++++-{}+-{}+-{}++-{}++++++-{}-{}+-{}
\end{flushleft}}

\para{Design 16.}
{\footnotesize\ttfamily
\begin{flushleft}
A = -{}-{}++-{}+-{}+-{}-{}+-{}++-{}-{}-{}++-{}++-{}-{}-{}-{}+-{}++++-{}\\
\hspace*{2.6em}++-{}-{}++-{}+-{}+++-{}-{}-{}+-{}+-{}-{}+-{}-{}-{}++++-{}+-{}-{}+\\
\hspace*{2.6em}-{}-{}+++++-{}-{}-{}+-{}+-{}-{}-{}++-{}++++++++++++++\\[2pt]
B = +++-{}++-{}+-{}+-{}-{}-{}-{}+-{}++-{}+++++-{}+-{}-{}-{}+-{}+-{}\\
\hspace*{2.6em}-{}+-{}-{}++-{}-{}++-{}-{}+-{}+++-{}+++-{}-{}-{}++++-{}++++\\
\hspace*{2.6em}-{}-{}+++++-{}+-{}+++-{}-{}-{}++++++-{}-{}-{}+-{}-{}++-{}+-{}
\end{flushleft}}

\para{Design 17.}
{\footnotesize\ttfamily
\begin{flushleft}
A = -{}-{}+-{}++-{}+++-{}-{}+-{}-{}-{}-{}-{}+-{}++-{}-{}+-{}+-{}+-{}++-{}\\
\hspace*{2.6em}+++-{}++-{}-{}++++-{}+++++-{}++-{}-{}-{}++++++-{}-{}+\\
\hspace*{2.6em}+-{}+-{}-{}-{}-{}++-{}+-{}++++++-{}+++-{}-{}-{}-{}-{}++-{}+-{}+\\[2pt]
B = -{}-{}+++-{}++-{}-{}++-{}+-{}-{}+++-{}+++-{}-{}-{}+-{}+++++\\
\hspace*{2.6em}-{}++-{}-{}++++-{}-{}-{}-{}+++++-{}++-{}+-{}-{}+++-{}++++\\
\hspace*{2.6em}-{}+-{}+-{}-{}-{}-{}-{}+-{}-{}-{}+-{}+-{}-{}-{}+++-{}++-{}+-{}+++++
\end{flushleft}}

\para{Design 18.}
{\footnotesize\ttfamily
\begin{flushleft}
A = -{}-{}++-{}-{}-{}++++-{}+-{}+++-{}-{}-{}-{}++++-{}++-{}++++\\
\hspace*{2.6em}-{}++-{}-{}-{}++-{}+-{}+-{}+++++++-{}-{}-{}+-{}-{}-{}+++-{}-{}-{}\\
\hspace*{2.6em}-{}-{}-{}-{}+-{}++-{}-{}-{}+-{}++-{}++-{}+++-{}+-{}++++++-{}+\\[2pt]
B = -{}++++-{}-{}++-{}+-{}-{}++-{}-{}-{}-{}++-{}-{}+-{}+-{}++-{}+++\\
\hspace*{2.6em}++-{}-{}++-{}-{}++-{}-{}+++++-{}+-{}+-{}+++-{}+-{}-{}+-{}++\\
\hspace*{2.6em}-{}+-{}++++++++-{}+-{}+-{}++-{}-{}++-{}++-{}-{}-{}-{}-{}-{}++
\end{flushleft}}

\para{Verification.} Each pair was checked by code independent of the search. The agreements satisfy
$P_A(s)+P_B(s)=2$ at all 98 nonzero shifts, the $198\times198$ matrix satisfies $MM^{\top}=(2v-2)I+2(J\oplus J)$, and
$\log\det M$ equals the Ehlich--Cohn bound $\log(2n-2)+\frac{n-2}{2}\log(n-2)$ for $n=198$ to machine precision. The
search's own exact recheck agreed, and every design was verified again on a second machine.

\para{Inequivalence.} Two designs are equivalent if one can be turned into the other by symmetries that preserve
D-optimality. At the level of sequences these are cyclic shifts, negations, multiplication of positions by a number
coprime to 99, and swapping $A$ and $B$. No combination of them maps any of the nineteen pairs (the published
design~\cite{r27} and designs 1--18) to another. At the level of matrices we compare 4-profiles. A 4-profile is the
multiset of $|\langle r_i\circ r_j, r_k\circ r_l\rangle|$ over all $\binom{198}{4}=62{,}117{,}055$ sets of four rows,
and it is unchanged by row and column permutations, sign changes and transposition. The largest value alone does not
separate every pair, since several designs share 68, 70, 74 or 86. The complete profiles differ for every pair,
however, and the number of 4-sets with value 2 already differs for all nineteen (Table~\ref{tab:profile}). The
nineteen matrices are therefore pairwise inequivalent.

\begin{table}[h]
  \caption{Two entries of each design's 4-profile, the largest value and the number of 4-sets of rows with value 2.
  The full profiles differ for every pair of designs.}
  \label{tab:profile}
  \small
  \begin{tabular}{@{}lrr@{}}
    \toprule
    Design & max $|\langle\cdot,\cdot\rangle|$ & 4-sets with value 2 \\
    \midrule
    published~\cite{r27} & 78 & 6,948,414 \\
    design 1 & 68 & 6,883,569 \\
    design 2 & 70 & 6,945,345 \\
    design 3 & 82 & 6,895,053 \\
    design 4 & 70 & 6,908,220 \\
    design 5 & 74 & 6,945,840 \\
    design 6 & 86 & 6,954,948 \\
    design 7 & 74 & 6,929,703 \\
    design 8 & 86 & 6,972,966 \\
    design 9 & 74 & 6,931,980 \\
    design 10 & 70 & 6,935,148 \\
    design 11 & 68 & 6,858,819 \\
    design 12 & 86 & 6,956,730 \\
    design 13 & 74 & 6,914,457 \\
    design 14 & 74 & 6,908,715 \\
    design 15 & 74 & 6,933,861 \\
    design 16 & 74 & 6,960,789 \\
    design 17 & 74 & 6,925,842 \\
    design 18 & 76 & 6,969,204 \\
    \bottomrule
  \end{tabular}
\end{table}

\para{End-to-end reproduction.} Design 1's job, re-run from its seed and started at the design's length-33 pair,
re-finds the design through the full pipeline of dealing, middle level, selection by join volume at the production
threshold, and exact test, with both the original and the faster filter settings. Design 4's job, re-run from its seed
on the server, reproduced the FABRIC result exactly, through the same parent and the same 1,097 units.

%% file: sections/B-lessons.tex
%% ---------------------------------------------------------------- Appendix B: lessons
\section{Lessons from an Unattended Exact Campaign}
\label{sec:eval-lessons}

The order-198 search of \S\ref{sec:eval-198} is exact: it tests every unit completely, so ``no design'' is a definite
answer for each unit tested. Running it unattended for weeks, first on one server and then across two testbed sites,
exposed five lessons for any deployment of this kind. Each lesson is stated first, followed by the evidence for it.

\para{Preemption must not waste work.} \emph{Record each unit as done when it finishes, and report finished work when
a job is stopped.} Jobs run under fixed wall-clock limits, 4 hours on the server and 2 hours on FABRIC, so that the
scheduler can rebalance work and deploy new versions. When progress was recorded only at the end of a long step, a job
stopped at its limit lost everything since its last record, and roughly half of all CPU time went to work that was done
but never counted. Recording each finished unit, checking the limit inside each step and reporting on termination
removed the loss. The same principle applies to pooled rounds. A failed round must leave no partially counted work.

\para{Memory sets concurrency.} \emph{Bound memory per job, in bytes, before counting cores.} At 2.9\,GB per process
the server could not run a job on every hardware thread. Storing each group of symmetric candidates once, in a compact
record, cut this to 0.65\,GB (Table~\ref{tab:checker}), after which memory rather than cores decided how many jobs
ran. A join table bounded only by its number of entries once reached 21\,GB, so every table needs a cap in bytes.

\para{Exactness needs an audit trail.} \emph{Make every counted unit traceable to the node, version and log line that
tested it.} Because each unit's completion was logged with its cache statistics, a cache bug found in code review
could be bounded after the fact to at most 210 of $9.0\cdot10^6$ units, instead of invalidating the run.

\para{Counting needs a known answer.} \emph{Check rate estimates against a known solution.} Before committing
thousands of CPU-hours, we projected the work per design at order 198 by measuring designs per unit at smaller lengths
($v=45$ and 63), where designs are plentiful, and extrapolating. Such a projection decides whether a search is
feasible and whether a long stretch without a design is bad luck or a bug. Each design appears in three units, as
three equivalent copies. It counts as one design, but units per design are therefore $3\times$ what a count of
design-holding units suggests. A known-answer test, which re-finds the published design and both of its copies,
exposed a projection that had missed this factor.

\para{The network's configuration is part of correctness.} \emph{Verify the network's addressing before every
experiment, not only at setup.} On FABRIC, an interrupted setup left three pairs of hosts sharing addresses, and a
lease renewal restored the duplicates after they had been removed. As a result, 138 of 376 runs could not connect and
had to be rerun. None produced a wrong answer, because every result is verified, but checking that each address has
exactly one holder would have prevented the loss.

%% file: sections/C-negative.tex
%% ---------------------------------------------------------------- Appendix C: negative results
\section{Negative and Calibration Results}
\label{app:negative}

This appendix collects the results that did not help, the limits we measured, and the corrections we made along the
way. Every idea was judged by verified solutions per CPU-second on held-out data. Most tests used small orders and
modest samples, so a negative result excludes only effects larger than its confidence interval at the tested length;
results that rest on proofs or exact counts are marked as such. The results are grouped by what they tell a network or
systems designer.

%% ------------------------------------------------------------------
\subsection{Today's search workloads give the network little to do}

\para{The measurement study.} We analyzed 63 network traffic traces of six workload classes: SAT clause sharing (14
portfolio runs and 2 MallobSat logs~\cite{r66}), LABS island search~\cite{r57}, Williamson meet-in-the-middle,
state-vector simulation, HPC (DUMPI~\cite{r52} and ATLAHS~\cite{r69}) and AI training (Llama 7B, and a mixture of
experts with expert parallelism 1, 4 and 8). Each class was tested against the three requirements of
\S\ref{sec:background}. The network must matter, meaning a 400\,Gb/s link speeds the run by at least $1.25\times$.
Reconfiguring must pay, meaning an oracle that reconfigures every 10\,\textmu s beats the best fixed topology by at least
15\%. And the solver must know something the network does not, meaning its telemetry improves forecasts of traffic
shifts over past traffic by at least 0.10 PR-AUC. No class met all three. SAT sharing used about 0.004\% of a link,
LABS migration moved 0.9--147\,KB per run, and the Williamson join's compute grew as fast as its bytes. The heavy
classes were stationary, and for all four AI traces a fixed topology chosen per job matched the clairvoyant oracle.
Telemetry improved forecasts by a mean of only 0.03 PR-AUC for LABS, and by 0.03 to 0.12 one round ahead for SAT.

\para{More sharing is not better search.} Clause buffers above 300 literals per worker per round did not help, and
heavier LABS migration lowered the hit rate. Forking whole island populations gave no gain (101 against 97 of 180
runs, $1.04\times$, 95\% CI 0.86--1.26). Replicating every Ramsey canonical form cost 0.35--9.7\,GB and saved nothing.
Sharing records of parents already shown to hold no solution would also save little. Only 0.1--2.3\% of 9,930 logged
Hadamard parents recurred across 33 runs, and parents that failed in one run and were retried in another were as
productive as fresh ones ($1.11\times$, 0.98--1.25).

\para{Compute-bound exchanges.} On FABRIC, when compute dominates the round, the network made no difference. At
$v=96$ and 16 hosts, link rates from 0.05 to 25\,Gb/s gave 62--66\,s to a first solution. Packet, demand-driven and
rotor-like exchange schedules, with gaps from $10^{-5}$ to $10^{-2}$\,s, gave 13--15\,s at $v=88$. Packed 12-byte
records against raw 64-bit keys, and a barrier against joining on arrival, gave the same results at 0.05--1\,Gb/s.
Captured traffic shows why. The exchange is a few short bursts per round, far below link rate.

%% ------------------------------------------------------------------
\subsection{Luck cannot be exploited}

\para{No lucky runs.} Following~\cite{r34,r40}, we traced about 5,500 Hadamard tower runs at $v=64$--80 to see whether
early behaviour predicts fast runs. At list sizes like those at the frontier, the wait to a solution is exponential
(coefficient of variation 0.96--1.00). Parents are interchangeable at $v=64$--72 when lists are not saturated. Success
per attempt is constant across attempts (hazard ratio 1.04, 95\% CI 0.93--1.16, over 2,500 towers), and parent features
give AUC 0.49--0.53, where 0.5 is chance. Early-run predictors score 43--55\% against a 50\% baseline, and dynamic or
fixed restart policies give 0.5--1.0$\times$ on held-out runs, including when trained at $v=64$--72 and applied at 80.
In quasigroup completion such prediction reached 60--96\% accuracy, but in the tower luck is pure chance. The useful
corollary is that a parent should be kept across rounds rather than discarded ($2.6\times$ at $v=64$ with 8 join rounds
per parent).

\para{Starting points are interchangeable.} Solutions per compressed pair have a dispersion index of 1.00--1.10, and the
richest 10\% of starting points hold the share a uniform model predicts, so choosing where to enter the tower cannot
help. The same holds for coarse parents. Over 111 distinct parents at $v=63$ the dispersion index is 0.92 (95\% CI
0.73--1.12), the richest 10\% hold 14.3\% of solutions against 14.4\% for a uniform model, and none of nine cheap parent
features ranks parents usefully. Parent features also did not predict Hadamard hits at $v=72$ (61 parents, 30,122
rounds, AUC 0.49--0.50), and choosing the best quarter of parents gains at most $1.16\times$. Whether D-optimal parents
extend to a solution is weakly predictable (AUC 0.61--0.67), but not well enough to repay the test.

\para{Near-misses do not point to solutions.} We counted, in every unit at $v=63$ (5.1\,M units, 145 distinct
designs), the pairs whose agreement sums miss the target by at most 4--12. Near-miss counts predict which units hold
solutions only through unit size (AUC 0.65 for both). Solutions do not cluster by compressed pair (dispersion 1.05,
chance band 0.83--1.19) or by parent, and visiting near-miss hot spots first gains 0.65--1.10. The one structure found
is reuse of length-9 compressions across designs (a sequence's compression is $2.2\times$ richer once it has hosted a
design), which helps only after designs are found. This matches the ``golf course'' picture of low-autocorrelation
sequences, whose optima are isolated~\cite{r11}.

\para{Do order-198 designs cluster?} After designs 3 and 4 were found, we asked whether parents that had yielded a
design hold more. The 16-host FABRIC site searched the four parents of designs 1--4 completely. It re-found each of the
four designs as its three equivalent copies and found nothing else, consistent with the parent-level homogeneity
measured at $v=63$. The fourteen designs found afterwards came from ordinary dealt jobs on other parents. Whether
designs concentrate in a coarser sense, for example in parents with high join volume (\S\ref{sec:eval-198}), remains
open.

%% ------------------------------------------------------------------
\subsection{Ideas that did not speed up the search}

\para{Search ideas.} The following gave no gain.
\begin{itemize}
  \item Exchanging key histograms or sketches before pooling. Volumes on one parent differ between machines by under
  1\%, and volume across parents is zero.
  \item Gossiping Bloom filters of barren subtrees. There were 0 repeats in 33,698 parent entries at $v=64$, and where
  repeats are common, skipped subtrees held their share of solutions.
  \item Local search seeded by a relaxation, with the same success per restart and $1.3$--$1.4\times$ slower.
  \item Lists built by recombination ($0.90\times$), since generation is only 3--6\% of CPU.
  \item Distinguished-point collision search under a memory cap ($0.04$--$0.18\times$). This was measured at D-optimal
  $v=45$, where memory does not bind, so it is not evidence against the method at the memory-bound frontier.
  \item Allocating work by branching dispersion (offspring dispersion at most 1.1).
  \item Enumerating cyclotomic norm equations, which repeats the compression levels or overshoots realizable sequences
  by $10^6\times$.
  \item Topological features, which vanish once matched on vector length.
  \item A Fourier-domain join, which provably cannot help, since an invertible change of keys preserves the number of
  matches.
\end{itemize}
Non-cyclic groups gave no gain at $v=63$ and are untested at the target. Their D-optimal solution density is
0.32--4.06$\times$ that of cyclic groups with no trend in $v$, and per CPU they gave $1.28\times$ at $v=45$ and
$1.01\times$ (0.90--1.14) at $v=63$. Pruning the order-198 exact test by partial spectral energy gave a modest
$1.14\times$ at $v=63$ while keeping 93\% of solutions, but coverage must then be counted at the kept fraction. Across
constructions, solution density per unit of search differs by orders of magnitude (Goethals--Seidel $\gg$ propus $>$
Williamson, by 8--150$\times$), so density is a property of the group and template, not of search order.

\para{Tower refinements.} A sum-of-squares penalty at coarse levels slowed the Hadamard tower ($v=64$, 5 of 8 solved,
median 221\,s). A local-move candidate generator for D-optimal kept only 3.7\% of moves feasible, produced 62--72\%
duplicates and was 1.9--2.1$\times$ slower. Screening coarse parents with a cheap trial round gave 0.88--0.96$\times$,
and successive halving gave 0.91--1.04$\times$ at $v=51$ and 57 (300 seeds per arm, enough to detect a $1.25\times$
effect). An earlier $1.75\times$ from 8 seeds was noise. Partial-match scores predict which parents extend to a
solution, but only weakly (AUC 0.61--0.67 on 200 parents at $v=57$, about as well as list size), too weakly to repay
screening. The fast candidate generator gave nothing for Hadamard, where 67--72\% of candidates pass the filter and the
join takes 60--80\% of the final level. Lists larger than a parent's distinct completions waste time (90\% duplicates
at $N=16$\,K and $v=51$, 7.7\,s against 1.0\,s). Two coprime folds, which made the Hadamard middle level about
$350\times$ faster (\S\ref{sec:eval-solvable}), failed when applied at the final level instead (no unit extended to a
solution in 25 minutes at $v=96$), because final-level units rarely do.

\para{Order-198 refinements.} Caching block lists (97.7--99.9\% hit rate), a hash join in place of sorting, a sorted
index on one coordinate, bounds on runs of blocks and frequency ordering each changed run time by a few percent at
most. Only the careful filter of \S\ref{sec:design} mattered. Restricting the search to designs fixed by a multiplier
subgroup also failed. At $v=99$, all 17 subgroups of order at least 4 were searched exhaustively and contain no
designs, and the analogous cases at $v=45$ contain none either. The restricted search is sound, since it found 1,190
verified symmetric designs at $v=33$, 57 and 63. But symmetric designs are rare. None of about 500 designs found by the
general method at $v\ge33$ is symmetric, so the restriction defines a separate search, not a faster one. The four
smallest subgroups (orders 2 and 3) leave $10^{14}$--$10^{15}$ candidates per sequence and were not searched.

\para{Screens for empty units.} An exact two-sided bound on one sequence's spectrum, taken from the other sequence's
survivors, is sound but rejects almost nothing at $v=99$ (99.8\% of survivors kept, 4.6\% slower), because the
survivors span the full range at every frequency. Ranking units by survivor counts, measured or estimated from a
sample, gave a net 0.97--0.99$\times$ at $v=63$, and even an oracle ranking gave only $1.04\times$. Unit cost and
solutions are uncorrelated ($r=0.000$ over $1.5\cdot10^7$ symmetry groups at $v=63$). Skipping costly units with
hindsight could gain 1.4--1.5$\times$ at $v=99$ (33 pairs), but no predictor available before the exact test captures
it (0.80--0.98$\times$). A 16-bit version of the filter was exact but 8--11\% slower.

\para{Reusing work across units.} Testing one sequence depends only on its own compressions, so the result could serve
every unit that shares them. At $v=99$ such sharing is rare. Each length-33 compression appears in 1.01--1.13
compressed pairs, against a guess of about 8 before measurement, so reuse gains at most about $1.1\times$ over uniform
sampling. Restricting sampling to shared compressions gave $1.40\times$ more units per CPU-hour, but those units hold
0.86--0.89$\times$ as many solutions, a net gain of about $1.2\times$, with identical results on all compared units. For
a distributed deployment this is a placement hint (co-locate units that share a compression), not a new kind of
exchange.

\para{Joining across parents.} Joining half-pairs across coarse parents is sound. Exact counts at $v=32$ give a hit
probability per pair of $8.4\cdot10^{-6}$ across parents against $8.2\cdot10^{-6}$ within one. But random parents
almost never share a coarse class (2,978 classes for 3,000 parents at $v=32$, and all distinct at 40 and 48), so
joining across parents amounts to a longer list on one parent and gains nothing. Pooling within a parent across nodes
remains the useful exchange.

\para{Projection methods.} RRR and Douglas--Rachford~\cite{r7,r23}, used in place of local search at the Hadamard
middle level, found exact solutions (verified order-260 matrices at $v=64$) but took 4.6--8.6\,s per solution against
0.053\,s, and found none in 60\,s at $v=72$--96, where local search needs 0.33--9\,s. Thirty settings of $\beta$ and
attempt length gave the same rate. Kline's relaxation~\cite{r43}, another continuous method, stops scaling at about
length 30 for four circulant blocks. Its recovery theorem, that an $n\times n$ Hadamard matrix is determined by all but
$O(n^2/\log n)$ random entries, relies on the $n^2$ redundant entries of an unstructured matrix, whereas a circulant
search has only $4v$ free signs, so the theorem does not carry over. Split across nodes (divide and concur), each
iteration would need two all-to-all exchanges of about 1.5\,KB against 0.5--4\,\textmu s of compute, so it is
latency-bound and offers no useful traffic for a fabric.

%% ------------------------------------------------------------------
\subsection{One positive result: sampling where solutions live}

Solution components are spectrally flatter than random feasible completions of the same parent
(Kolmogorov--Smirnov $p<10^{-20}$). Keeping only the flattest 30--50\% of completions (``tilting'') gave
5.3--10.3$\times$ more verified hits per CPU in isolated joins at $v=72$, sampler included. In full tower runs, whose
list sizes already succeed in most rounds, the gain appears as earlier first solutions. Time to the first verified
Hadamard solution fell $2.0\times$ (1.6--2.5) at $v=80$ and about $3\times$ (1.8--4.1, six seeds) at $v=96$. At four
times the list size, as in a pooled round, the gain vanished (0.9--1.2$\times$), and a quarter or more of tilted lists
could not fill, because tilting shrinks the completion space that bounds a pool. Tilting therefore helps independent
nodes but competes with pooling for the same parent's completions. Combining the two means choosing the tilt and the
pool size together.

%% ------------------------------------------------------------------
\subsection{Limits}

\para{An open order.} A pilot for D-optimal $v=135=3^3\cdot5$ (order 270, open), whose only row sums are $(3,23)$,
found final exact tests as cheap as at $v=99$, but with about 100 survivors per side instead of thousands and 36
coprime frequency pairs to match. The estimated success per unit is about $10^{-14}$, against $4\cdot10^{-8}$ at
$v=99$, and the cost to a first design about $10^{10}$ thread-hours (range $10^9$--$6\cdot10^{11}$) on today's cores.
Exhaustive search of designs fixed by multiplier subgroups of order 8--72 found none.

\para{Where the levels fail.} Coarse-level matching costs time on short coarse problems (Hadamard $v=38$, 2.5\,s against
0.16\,s, and D-optimal $v=57$, 4.0\,s against 2.9\,s), and a single fold with a coarse join is slower than a deeper fold
chain where one exists ($v=63$, 114\,s against 24\,s). Hadamard fold-5 steps placed last failed ($v=60$, 0 of 3), and
fold-5 steps with large entries ($M\ge20$, forced at $v=500$) gave 0 hits in 343 join rounds and 0 of 23,837
local-search attempts. At D-optimal $v=75$ every single-chain variant failed (0 of 3), including one with a mid-level
join. $v=87$ needs about 4\,M coarse candidates per sequence (55\,s per coarse solution) and was not attempted with a
single chain. Sampling inside double-compression pieces failed (30\,s per round, 4--8\,K distinct candidates, no hits),
and with $p=3$ every pair of coarse projections was compatible, so the second projection helps only by making the
pieces small enough to enumerate exactly.

%% ------------------------------------------------------------------
\subsection{Corrections made during the study}

A LABS task-dealing bug had inflated load imbalance, which is now 1.2--1.7$\times$. A weak Ramsey baseline overstated
gains of 4.1--13.7$\times$, which are now 0.97--2.3$\times$. An incremental-update error at shift $v/2$ in Hadamard
local search was fixed. The Hadamard hit exponent was re-measured with all hits counted ($N^{3.8\text{--}4.1}$ rather
than $N^{3.3}$), which replaced an earlier pooled advantage of $P^{0.66}$ with about $P$. Extrapolations over 80 units
of length are order-of-magnitude estimates only.

%% file: sections/D-theory.tex
%% ---------------------------------------------------------------- Appendix C: theory
\section{Theory of Towers and Joins}
\label{app:theory}

\begin{figure}[t]
  \centering
  \includegraphics[width=\columnwidth]{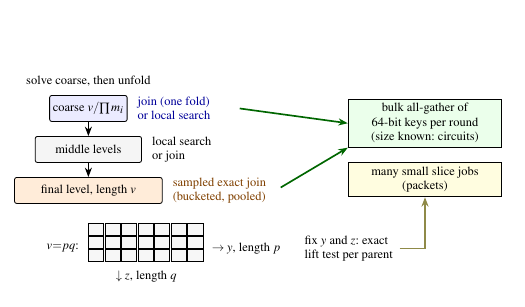}
  \caption{The full partitioned search (Figure~\ref{fig:tower} gives a toy example). A compression tower splits a
  solution into exact coarse-to-fine levels; each level uses local search or a sampled exact join, chosen by where time
  goes. Joins exchange compact key lists in rounds of known size; double compression ($v=pq$) turns the final level into
  small independent slice jobs.}
  \label{fig:pipeline}
\end{figure}

We write $x:\Z_v\to\{\pm1\}$, $\omega_v=e^{2\pi i/v}$,
$X(k)=\sum_i x_i\omega_v^{ik}$ and $\PAF_x(s)=\sum_i x_ix_{i+s}$. A
problem asks for $r$ sequences with $\sum_j\PAF_{x_j}(s)=c$
for all $s\ne0$ ($r=4$, $c=-4$ for Hadamard; $r=2$, $c=2$ for
D-optimal). Taking the DFT, this is equivalent to $\sum_j|X_j(k)|^2=rv-c$ for all $k\ne0$ ($4v+4$ and $2v-2$). We separate
what is proved (T1, Lemma~\ref{lem:ceiling}, Proposition~\ref{prop:invariance}), what holds in an idealized key model (T2,
T3), what is conditional on a conjecture (the 4SUM remark), and what is a formulation whose parameters must be measured
(the tower DP).

\subsection*{T1. Folds are subgroup quotients}

For $m\mid v$ let $H_m=\langle v/m\rangle\le\Z_v$, the subgroup of order $m$, and $\pi:\Z_v\to\Z_v/H_m\cong\Z_{v/m}$,
$i\mapsto i\bmod v/m$. The fold by $m$ is the pushforward $y=\pi_*x$, $y_j=\sum_{\pi(i)=j}x_i=\sum_{t<m}x_{j+tv/m}$.

\begin{proposition}\label{prop:fold}
(a) $Y(k')=X(mk')$ for all $k'\in\Z_{v/m}$. Hence two sequences have the same fold if and only if their DFTs agree on
$H_m^{\perp}=m\Z_v$, the $v/m$ frequencies whose characters are trivial on $H_m$. (b) $\PAF_y(j)=\sum_{\pi(s)=j}\PAF_x(s)$.
Hence if $x_1,\dots,x_r$ satisfy the PAF condition with constant $c$, their folds satisfy it exactly with constant $mc$
(targets $-4m$ and $2m$).
\end{proposition}
\begin{proof}
(a) $X(mk')=\sum_ix_i\omega_v^{imk'}=\sum_ix_i\omega_{v/m}^{ik'}$, and the exponent depends only on $\pi(i)$; grouping by
fibres gives $\sum_jy_j\omega_{v/m}^{jk'}=Y(k')$. The DFT on $\Z_{v/m}$ is invertible, so $y$ and $X|_{m\Z_v}$ determine
each other. The character $i\mapsto\omega_v^{ik}$ is trivial on $H_m$ iff $\omega_v^{kv/m}=1$ iff $m\mid k$. (b) In the
group algebra $\mathbb{C}[\Z_v]$, $\PAF_x=x*\tilde x$ with $\tilde x(i)=x(-i)$. Pushforward along a homomorphism is a ring
homomorphism that commutes with $\tilde{\ }$, so $\PAF_y=\pi_*\PAF_x$. For $j\ne0$ the fibre $\pi^{-1}(j)$ contains $m$
nonzero shifts, so the sum over sequences is $mc$.
\end{proof}

\para{Consequences.} (i) \emph{Canonicity.} Call a compression exact if it is the pushforward along a surjective
homomorphism $\varphi:\Z_v\to Q$. Since $Q\cong\Z_v/\ker\varphi$, exact compressions correspond one to one to subgroups,
for $\Z_v$ to divisors of $v$; Proposition~\ref{prop:fold} holds verbatim for any finite abelian group with $H^{\perp}$ the
annihilator in the dual group, which covers the non-cyclic groups of Appendix~\ref{app:negative}. (ii) \emph{Towers are
chains.} Folding by $m$ and then by $m'$ is folding by $mm'$, because $H_m\le H_{mm'}$. A tower is a chain
$\{0\}<H_{m_1}<H_{m_1m_2}<\cdots$ in the subgroup lattice, and since $H\le K$ implies $K^{\perp}\le H^{\perp}$, unfolding
from $K$ to $H$ adds exactly the frequencies $H^{\perp}\setminus K^{\perp}$: the tower reveals the spectrum block by block,
and each frequency's identity $\sum_j|X_j(k)|^2=rv-c$ is enforced exactly at the coarsest level that sees it. (iii)
\emph{Depth limit.} $k\in H^{\perp}$ for some $H\ne\{0\}$ iff $\gcd(k,v)>1$, so the $\phi(v)$ frequencies coprime to $v$
are seen only by the final level, whatever the tower (60 at $v=99$, the 60 frequencies of the order-198 lift). (iv)
\emph{Double compression.} Folds by distinct primes $p$ and $q$ dividing $v$ are an antichain, not a chain; folding by
every prime divisor $p_1,\dots,p_r$ of $v$ together sees $\bigcup_ip_i\Z_v$, which is every $k$ with $\gcd(k,v)>1$. When
$v$ has two prime factors ($99=3^2\cdot11$: folds by 3 and 11; $96=2^5\cdot3$: folds by 2 and 3), two folds therefore
already carry all the exact information any tower can, which is why double compression turns a parent into small exact
units. The natural design space is thus the subgroup lattice, with chains and antichains as moves.

\para{Choosing a tower (formulation).} Treat each level as a sequential test whose order is constrained to the lattice.
For an unfold edge $K\to H$ ($H<K$), let $c(K,H)$ be the expected cost of one lift attempt on a $K$-level parent and
$s(K,H)$ its probability of yielding an $H$-level solution; for a join edge both depend on the list size, and $c$ is the
minimum over $W$ of T2's cost with $s=1-e^{-\lambda(W)}$. Under the memorylessness we measured (independent attempts,
interchangeable parents; Appendix~\ref{app:negative}), the expected cost to a first $H$-level solution is
\[
C^*(H)=\min\Bigl\{E(H),\ \min_{K>H}\frac{C^*(K)+c(K,H)}{s(K,H)}\Bigr\},
\]
with $E(H)$ the cost of solving level $H$ directly, and the answer is $C^*(\{0\})$. Because $C\mapsto(C+c)/s$ is
increasing, Bellman's principle applies and the recursion is solved exactly over the $d(v)$ lattice nodes (12 at $v=96$)
once the edge parameters are known; parent reuse divides $C^*(K)$ by the attempts per parent. This is a formulation, not
a theorem about specific orders: the parameters must be measured, and it ignores parent-fertility variation, which
matters when most parents are barren (\S\ref{sec:discussion}).

\subsection*{T2. Which join, at which cost ratio}

\para{Generic-key model.} Each candidate's key is uniform on an abelian group $G$ of size $S$ and independent of all other
keys (the random-oracle idealization of the linear hash), and a $k$-tuple with one candidate per list is a solution iff
its keys sum to a target. Real keys deviate (three low bits are constant on real lists; solution density is non-uniform,
T3); those deviations are exactly where gains beyond T3's ceiling come from.

\para{Counts.} With $W$ candidates per list the expected number of solutions is $W^2/S$ for two lists and $W^4/S$ for four
(Lemma~\ref{lem:ceiling}). A four-list join that forms pair records $h_a+h_b$ on one side and matches them against
$h_c+h_d$ on the other, in any partition into key classes, finds in expectation $W^2/S$ solutions per $AB$ record, since
each record completes with $W^2$ $CD$ pairs, each a solution with probability $1/S$; it therefore enumerates $2S/W^2$
records per solution, whatever the partition. The full bucketed join uses all classes: $2W^2$ records, all $W^4/S$
solutions, so it attains the ceiling per candidate. Wagner's merge on $\ell=\log_2W$ bits uses one class: $2W$ records
and $W^3/S$ solutions. The measured exponents match: full-join hits $\propto N^{4.11}$ and $N^{3.96}$ with $N^{2.00}$
records, Wagner $N^{3.02}$ with $N^{1.00}$ work (4-list XOR).

\para{Conditional optimality.} The full join runs in $O(W^2+\text{solutions})$ time. Deciding in the worst case whether
four lists of $W$ keys contain a solution in $O(W^{2-\varepsilon})$ time would refute the $k$-SUM conjecture for $k=4$,
which posits $W^{\lceil k/2\rceil-o(1)}$ and is proved in the $k$-linear decision-tree model~\cite{r1,r24,r58}. This is a
worst-case statement: random keys are exactly where Wagner finds a subset of the solutions faster, so it shows only that
the $W^2$ per round of a complete join cannot be improved in general. Schroeppel--Shamir~\cite{r67} enumerates the same
pair sums in sorted order with $O(W)$ memory; it changes memory, not pair counts.

\para{Two-parameter cost.} Let cost $=c_q\cdot(\text{candidates})+c_p\cdot(\text{pair records})$. With four lists of $W$ and
a round that stops at its first hit, classes are enumerated in random order and the expected cost to a first solution is
\[
C(W)\approx4c_qW+2c_pS/W^2\qquad(W\ge S^{1/4}),
\]
ignoring per-class overhead (at most the record count when $2^{\ell}\le W$). It is minimized at $W^*=(c_pS/c_q)^{1/3}$ with
$C^*=6c_q^{2/3}(c_pS)^{1/3}$, where candidates take two thirds of the cost; this is the two-thirds overhead share and the
$\kappa^{2/3}$ scaling found by the network model (\S\ref{sec:eval-network}). Wagner's single class at $W=S^{1/3}$ costs
$(4c_q+2c_p)S^{1/3}$, so with $\rho=c_q/c_p$
\[
C_{\mathrm{Wagner}}/C^*=(2\rho+1)/(3\rho^{2/3})\ge1,
\]
with equality iff $\rho=1$. Expensive candidates ($\rho>1$) call for shorter lists and many classes, i.e.\ the bucketed
full join with early stop; cheap candidates ($\rho<1$) call for longer lists and fewer classes, the Wagner direction. If
memory or a parent's finite completion pool caps $W$ below $W^*$, every class should be joined: at equal $W$ the full
join's cost per solution $S(4c_q/W^3+2c_p/W^2)$ is below Wagner's $S(4c_q+2c_p)/W^2$. On one node our Hadamard lists are
expensive (list generation is 93--97\% of Wagner's time) and pool-capped (smallest pool 2.5--17$\Nstar$ at $v=96$), so
$\rho\gg1$; measured, Wagner was 3--15$\times$ slower class by class and 4--46$\times$ slower as a single class.

\para{Pooling changes the effective costs.} On $P$ nodes the exchange decides which terms shrink with $P$. With the key
all-gather each node generates $4W/P$ candidates but receives and indexes all $4W$ keys, so per node
$c_q^{\mathrm{eff}}=c_q/P+c_g$ and $c_p^{\mathrm{eff}}=c_p/P$. With a residue (Wagner-style) shuffle, candidates and pair
records each cross the network once, so $c_q^{\mathrm{eff}}=(c_q+c_s)/P$ and $c_p^{\mathrm{eff}}=(c_p+c'_s)/P$.
Substituting into $C^*$, per-node time to a first solution is
\begin{align*}
T_{\mathrm{AG}}&=6\Bigl(\frac{c_q}{P}+c_g\Bigr)^{2/3}\Bigl(\frac{c_pS}{P}\Bigr)^{1/3},\\
T_{\mathrm{SH}}&=6(c_q+c_s)^{2/3}(c_p+c'_s)^{1/3}S^{1/3}/P.
\end{align*}
$T_{\mathrm{AG}}$ falls as $P^{-1}$ only while $Pc_g\ll c_q$ and then as $P^{-1/3}$, while $T_{\mathrm{SH}}$ falls as
$P^{-1}$ throughout; the shuffle wins beyond $P^*\approx[(c_q+c_s)\sqrt{1+c'_s/c_p}-c_q]/c_g$, and a cheaper generator
(smaller $c_q$) lowers $P^*$ whenever $c'_s>0$. The all-gather also holds $4W$ keys per node against $4W/P$ for the
shuffle, so under a memory cap only the shuffle lets $W$ grow with $P$. The optimal join is therefore a function of $P$.
This matches our measurements and measurement-parameterized projections: on one node the full join wins; pooled Wagner is
0.52--1.02$\times$ the best full-join scheme at $P\le256$, 1.16--1.23$\times$ faster at $P=1{,}024$ on announced circuits
(0.86--0.92$\times$ on a 3:1 leaf-spine), and 1.9--4.1$\times$ faster at $P=8{,}192$ with 5--37$\times$ fewer bytes per
node; a 4.6$\times$ faster generator moves the crossover down to $P\approx256$; and with memory charged only the shuffle
reaches $v\approx161$--177 on 1,024 GPUs (\S\ref{sec:eval-network}). The model predicts the direction and the existence
of a crossover, not its exact location, which depends on constants it omits (sorting, per-class overhead, fabric).

\subsection*{T3. The luck ceiling}

\begin{lemma}[Luck ceiling]\label{lem:ceiling}
In the generic-key model, let an algorithm evaluate candidates adaptively (each evaluation reveals a fresh uniform key)
and stop at any stopping time, having evaluated $W_i$ candidates in list $i$. Then the expected number of solutions among
evaluated $k$-tuples is $\E[\prod_iW_i]/S$. For fixed $W$ per list this is exactly $W^2/S$ ($k=2$) and $W^4/S$ ($k=4$);
for a budget of $W$ evaluations in total it is at most $(W/k)^k/S$.
\end{lemma}
\begin{proof}
Fix a tuple $\tau$. Whether $\tau$ is completed is decided before the key of its last-evaluated member is revealed, and
that key is uniform and independent of everything before it, so
\[\Pr[\tau\text{ completed and a solution}]=\Pr[\tau\text{ completed}]/S.\] Sum over $\tau$ (linearity of expectation, monotone convergence) and bound $\prod_iW_i$ by AM--GM.
\end{proof}

For four lists the ceiling on finding is the per-record count of T2: at most $W^2/S$ expected solutions per $AB$ pair
record, for records formed without regard to the $CD$ keys, as in every class-based join. Restarts and choice of starting
point cannot raise either bound, because the bound depends only on how many tuples are formed, which is the formal
content of the memorylessness we measured. Solutions per unit cost are at most (tuples per unit cost)$/S$, so a gain must
come from one of four sources: (a) more tuples per candidate through larger shared lists (pooling, $W^{k-1}$ per
candidate); (b) cheaper candidates or cheaper rejection, as with the mixed-precision filter (3.83$\times$ on identical
work) and partial-energy pruning (1.14$\times$ at $v=63$, keeping 93\% of solutions); (c) spending work where tuples per
cost are higher, as join-volume selection does (6.1--6.3$\times$ measured on the 64-thread server, \S\ref{sec:eval-198}),
with solutions proportional to pair volume (slope 0.96--0.97), i.e.\ density per pair flat as the model predicts; or (d)
departing from the model through non-uniform solution density, as tilted sampling of spectrally flatter completions does
(5.3--10.3$\times$ at $v=72$, validation at larger $v$ pending).

\begin{proposition}[Invariance]\label{prop:invariance}
If $T:G\to G'$ is an injective homomorphism, then $\sum_ik_i=t$ iff $\sum_iT(k_i)=T(t)$, so the solution set, the
pair-record counts and the hits of every join are unchanged; only the cost per record can change.
\end{proposition}

The proof is immediate. The Fourier-domain join replaces PAF vectors by power spectra, an invertible linear map (the
DFT), so it finds exactly the same tuples, and our negative result for it (Appendix~\ref{app:negative}) is a theorem
rather than a measurement. A non-injective map such as the 64-bit hash can only add false matches, which the exact
recheck removes.

\para{Alternative search structures.} We tested five alternatives on the same parents and lists as the tower; every
reported solution was verified exactly (2,953 Hadamard matrices), and none beats the current design. \emph{Compression
lattice.} Adding one coprime (CRT) compression directly above the final join, using the smallest odd divisor, is best:
1.37$\times$ the pure fold chain at $v=72$ and 1.61$\times$ at $v=60$ (230 seeds); at $v=84$ the chain alone never solved.
This is the structure our kits already use. Larger divisors, CRT nodes at two levels, or a node at the final level give
0.13--0.73$\times$; a final-level node cuts memory from 520\,MB to 12--34\,MB but costs 5--8$\times$ the CPU. \emph{Joins
under a memory cap.} At caps from 100\% down to 0.1\% of the full table, the bucketed join wins at every cap and is
fastest at the smallest (cache-sized buckets). Relative solutions per CPU: Schroeppel--Shamir~\cite{r67}
0.09--0.15$\times$, Wagner by class 0.18--0.56$\times$, classic Wagner about 1/600. A memory cap therefore costs a bucketed
join nothing; shrinking the bucket table alone makes the $v=72$ kit 1.55$\times$ faster (1.49--1.62) in 40\,MB instead of
520\,MB. \emph{Representation technique}~\cite{r35}. It reproduces the expected gain on subset sum (work $2^{0.43n}$ vs.\
$2^{0.50n}$) but finds no usable redundancy in PAF joins: PAF is quadratic in a sequence's entries, so sequences have no
additive split; using all three 2+2 pairings gives 0.91--0.92$\times$; and shift twins always share their key.
\emph{SAT+CAS} in the style of MathCheck~\cite{r12}, on Williamson sequences: exact, but 78--280$\times$ slower to a first
solution than our exact tower and 1,300$\times$ slower to all solutions at $n=33$; it did not finish $n=35$ or 39 in 30
minutes, where the tower needs 7--19\,s. \emph{Fold order.} A dynamic program over the subgroup lattice reproduces
measured costs within 2\% for 11 of 14 configurations and selects the orders already in use.

\para{Engineering the tower and the exchange.} We then tuned the single-node code and the exchange. Every faster variant
reproduces the baseline's output exactly on fixed seeds (same logs, survivor sets and solutions), and all 2,500+ reported
solutions were verified exactly; ratios are times to a first verified solution with bootstrap 95\% intervals. \emph{Join
enumeration.} Walking the buckets with a cursor instead of sorting each bucket gives 1.23$\times$ (1.20--1.27) at $v=72$,
1.49$\times$ (1.44--1.55) at $v=80$ and 1.22$\times$ (1.18--1.27) at $v=88$ over the previous code at its own best table
size, and needs 15--188\,MB instead of 0.4--0.7\,GB. The time per examined pair falls from 18--20 to 12--13\,ns. The best
bucket table holds about 32--80 times the list size $N$: smaller tables pay a per-bucket cursor cost, larger ones overflow
the L2 cache. \emph{Candidate generation.} Computing each candidate's key exactly from a quadratic form in its free signs
makes generation 8.3$\times$ cheaper per candidate. End to end that is 1.65$\times$ at $v=72$, 1.20$\times$ at $v=80$ and
1.04$\times$ at $v=88$, where the join dominates. \emph{Stacked with tilt.} The two gains multiply with tilt:
4.3--4.5$\times$ over the previous kit at $v=80$, and about 6.5$\times$ at $v=88$ and 5.5$\times$ at $v=96$ (wide
intervals; few seeds). \emph{Tilt versus pooling.} In the hash-partitioned emulation a tilted pool beats an untilted pool
of the same size only below saturation (1.8$\times$ at $v=72$, 2.1--2.7$\times$ at $v=80$); in large pools it loses
(0.3--0.8$\times$), because tilt and pooling compete for a parent's completions. \emph{Order-198 checker.} Running all 30
frequencies in the single-precision filter gives 1.34$\times$ (1.30--1.40) at $v=99$. Hand-written AVX-512 adds 4\%,
which we do not expect to carry over to CPUs with one 512-bit FMA unit per core; automatic 512-bit vectorisation
(0.90$\times$), frequency reordering (no gain) and prefetching with chunked early exit (0.79$\times$) do not help.
\emph{Wire format.} Sorting each block of keys and coding it with Elias--Fano, with ids left implicit, halves the
all-gather from 12 to about 6\,B per candidate; truncated keys add false matches at the expected rate $N^4/2^t$. In the
network model this gives 1.4--1.55$\times$ at $P\ge256$ on GPU and accelerated nodes, and at most 1.3$\times$ on CPU nodes
(on FABRIC's CPU hosts the two formats gave the same results, \S\ref{sec:fabric}). Sending each candidate's free signs
instead (about 2\,B) and recomputing its key on the receiver (6--21\,ns per key with byte-block tables) adds
1.25--1.6$\times$ on accelerated nodes, but only when the recompute runs on the accelerator. \emph{Overlap.} In real
multi-process runs over TCP, overlapping the exchange with generation alone gives 1.0$\times$. Joining chunks as they
arrive hides about 83\% of the exchange: 1.2--1.66$\times$ per round, 1.2--1.4$\times$ in the model.
\emph{Final-join kernel (kit v5c).} Key-ordered list walks (1.4--2.6$\times$ per examined pair), compact 8-byte candidate
masks (4$\times$ larger lists in the same memory), skipping parents whose pool is exhausted, and 16 cooperating workers
per parent (two-process efficiency 0.95--0.99) together gave 5.3$\times$ (2.6--10.4) on one core at $v=104$
(\S\ref{sec:eval-solvable}).